\documentclass[twocolumn]{aastex6}
\usepackage{natbib}
\usepackage{color}
\usepackage{hyperref} %----set up hyper link to citations, equations...
\hypersetup{colorlinks=true,citecolor=blue}
\usepackage{times}
\usepackage[T1]{fontenc}
\usepackage{aecompl}
\usepackage{amsmath}
\usepackage{multirow}
\usepackage{threeparttable}
\usepackage{verbatim}

\def    \simlt  {\lower.5ex\hbox{$\; \buildrel < \over \sim \;$}}
\def    \simgt  {\lower.5ex\hbox{$\; \buildrel > \over \sim \;$}}

\newcommand{\rom}[1]{%
  \textup{\uppercase\expandafter{\romannumeral#1}}%
}

\def \bea {\begin{eqnarray}}
\def \ena {\end{eqnarray}}                  

\newcommand     \mum    {\,\mu{\rm m}}  % to use in math mode

\def	\cm		{\,{\rm {cm}}}

\def	\erg		{\,{\rm {erg}}}

\def   \exp 		{\,{\rm {exp}}}
\def	\g		{\,{\rm g}}

\def	\K		{\,{\rm K}}

\def	\s		{\,{\rm s}}

\def    \ln  		{\,{\rm {ln}}}

\def	\H		{\rm H}

\def    \aa  		{\rm {A\&A}}
\def    \apj  		{\rm {ApJ}}

\def    \mnras  		{\rm {MNRAS}}
\def    \araa  		{\rm {ARA\& A}}

\def    \Bv     	{\bf  B}

\def	\ba		{{\bf a}}

\def	\bJ		{{\bf J}}

\def	\gas		{\rm {gas}}

\def	\rad		{\rm {rad}}

\def    \ext    	{\rm {ext}}
\def    \pol    	{\rm {pol}}

\begin{document}
\shorttitle{Physical Model of Dust Polarization}
\shortauthors{Lee, Hoang, Ngan, and Cho}
\title{Physical Model of Dust Polarization by Radiative Torque Alignment and Disruption and Implications for grain internal structures}

\author{Hyeseung Lee\altaffilmark{1}, Thiem Hoang\altaffilmark{1,}\altaffilmark{2}, Ngan Le\altaffilmark{3,}\altaffilmark{4}, and Jungyeon Cho\altaffilmark{1,}\altaffilmark{5}}
\affil{$^1$ Korea Astronomy and Space Science Institute, Daejeon 34055, Republic of Korea; \href{mailto:thiemhoang@kasi.re.kr}{thiemhoang@kasi.re.kr}}
\affil{$^2$ Korea University of Science and Technology, 217 Gajeong-ro, Yuseong-gu, Daejeon, 34113, Republic of Korea}
\affil{$^3$ Institute of Astronomy, Faculty of Physics, Astronomy and Informatics, Nicolaus Copernicus University, Grudziadzka 5, 87-100 Torun, Poland}
\affil{$^4$ Department of Space and Applications, University of Science and Technology of Hanoi, Vietnam Academy of Science and Technology, 18 Hoang Quoc Viet, Hanoi, Vietnam}
\affil{$^5$ Chungnam National University, Daejeon 34134, Republic of Korea}

\begin{abstract}
Dust polarization depends on the physical and mechanical properties of dust, as well as the properties of local environments. To understand how dust polarization varies with grain mechanical properties and the local environment, in this paper, we model the wavelength-dependence polarization of starlight and polarized dust emission by aligned grains by simultaneously taking into account grain alignment and rotational disruption by radiative torques (RATs). We explore a wide range of the local radiation field and grain mechanical properties characterized by tensile strength. We find that the maximum polarization and the peak wavelength shift to shorter wavelengths as the radiation strength $U$ increases due to the enhanced alignment of small grains. Grain rotational disruption by RATs tends to decrease the optical-near infrared polarization but increases the ultraviolet polarization of starlight due to the conversion of large grains into smaller ones. In particular, we find that the submillimeter (submm) polarization degree at $850~\mu \rm m$ ($P_{850}$) does not increase monotonically with the radiation strength or grain temperature ($T_{d}$), but it depends on the tensile strength of grain materials. Our physical model of dust polarization can be tested with observations toward star-forming regions or molecular clouds irradiated by a nearby star, which have higher radiation intensity than the average interstellar radiation field. Finally, we compare our predictions of the $P_{850}-T_{d}$ relationship with {\it Planck} data and find that the observed decrease of $P_{850}$ with $T_{d}$ can be explained when grain disruption by RATs is accounted for, suggesting that interstellar grains unlikely to have a compact structure but perhaps a composite one. The variation of the submm polarization with U (or $T_{d}$) can provide a valuable constraint on the internal structures of cosmic dust. 
\end{abstract}

% ========== INTRODUCTION ==========

\section{Introduction}\label{sec:intro}
Dust is an intrinsic component of the interstellar medium (ISM) and plays important roles in astrophysics. Dust grains absorb and scatter starlight, and infrared emission from heated dust grains is a powerful probe of star and planet formation. Photoelectric effect from small dust grains is essential for heating and cooling of molecular gas, and grain surfaces are catalytic sites for molecule formation (see \citealt{Draine:2011}). 

The polarization of starlight (\citealt{1949Sci...109..166H}; \citealt{1949Natur.163..283H}) and polarized thermal emission (\citealt{1989IAUS..135..275H}) due to the alignment of dust grains with ambient magnetic fields allow us to map magnetic fields in various environment conditions, from the diffuse medium to molecular clouds to circumstellar regions (see \citealt{2003ApJ...598..392L}; \citealt{2012ARA&A..50...29C}; \citealt{2015ASSL...407.59}). Moreover, polarized thermal emission from aligned grains is a major foreground contamination of cosmic microwave background (CMB) that must be separated to accurately measure the CMB B-modes (\citealt{2003NewAR..47.1107L}). It is now established that an accurate model of dust polarization spectrum is required for the precise detection of B-modes (\citealt{2016A&A...586A.141P}). Such an accurate model of dust polarization depends on dust physical properties (size, shape, and composition), grain alignment with the magnetic fields, and the gas density and magnetic field structures.

The question of how dust grains become aligned with the magnetic field is a longstanding problem in astrophysics (see \citealt{2007JQSRT.106..225L} for a review). After seven decades of research, RAdiative Torque (RAT) alignment becomes the popular theory to explain grain alignment (see \citealt{2015ARA&A..53..501A} for a recent review). The idea of RATs was first introduced by \cite{1976Ap&SS..43..257D}, which were quantified based on the differential scattering and absorption of left- and right-handed photons (i.e., photon angular momentum). Numerical calculations for several realistically irregular shapes were carried out by \cite{1996ApJ...470..551D} and \cite{1997ApJ...480..633D}. However, an analytical model that provides physical insights into RATs and RAT alignment was formulated by \cite{2007MNRAS.378..910L} where RATs were quantified based on the transfer of photon momentum to the helical grain. Extended numerical calculations of RAT alignment for the different environments were carried out in \cite{2008MNRAS.388..117H}, \cite{2009ApJ...695.1457H}, and \cite{2014MNRAS.438..680H}. A unified theory of grain alignment for grains with iron inclusions is introduced in \cite{2016ApJ...831..159H}. Recently, numerical calculations of RATs for a large ensemble of grain shapes were presented by \cite{2019ApJ...878...96H}. 

One of the key predictions of RAT theory is that the degree of grain alignment depends on the local conditions, including the radiation field and gas properties (density and temperature). As a result, toward the center of a dense molecular cloud with low radiation intensity, only large grains can be aligned by attenuated interstellar photons (\citealt{2005ApJ...631..361C}). As a result, the peak wavelength of starlight polarization would increase with increasing visual extinction $A_{V}$ \citep{2015MNRAS.448.1178H}. This prediction was supported by observational data by \cite{2008ApJ...674..304W}. The angle-dependence of RAT alignment is then successfully tested by observations of starlight polarization by \cite{2011A&A...534A..19A}. Submm/FIR polarization of starless cores also reveals the existence of polarization hole (\citealt{2014A&A...569L...1A}; \citealt{2015ApJ...149.31J}), which is expected from the RAT theory. In the other regime of strong radiation sources, the RAT theory predicts an increased alignment of grains when the radiation strength increases, and the peak wavelength is shifted to smaller values. Such a prediction is consistent with observations toward type Ia supernovae (SNe Ia; \citealt{2017ApJ...836...13H}; \citealt{Giangetal:2019}). Therefore, the polarization degree of polarized emission is expected to increase with increasing radiation intensity according to the classical picture of the RAT theory (\citealt{2007MNRAS.378..910L}).

In addition to grain alignment, the grain size distribution is required for modeling dust emission and polarization spectrum. The grain size distribution is expected to evolve from the ISM to dense molecular clouds due to various physical effects. For instance, grains can be destroyed via grain shattering, thermal and non-thermal sputtering in interstellar shocks ( \citealt{1994ApJ...431..321T}; \citealt{1996ApJ...469..740J}). On the other hand, grains can grow in dense molecular clouds due to the accretion of gas species onto the grain surface as well as coagulation due to grain-grain collisions (see \citealt{2018ApJ...857...94Z} and reference therein). In the diffuse medium, grain shattering induced by grain acceleration by magnetohydrodynamic (MHD) turbulence (\citealt{2004ApJ...616..895Y}; \citealt{2012ApJ...747...54H}) is thought to determine the upper cutoff of the grain size distribution (\citealt{2009MNRAS.394.1061H}). However, a new physical mechanism, so-called RAdiative Torque Disruption (RATD), that dominates the upper cutoff of the grain size distribution, was recently discovered by \cite{2019NatAs...3..766H}. This RATD mechanism is based on the fact that suprathermally rotating grains spun-up by RATs induces centrifugal stress that can exceed the maximum tensile strength of grain materials, resulting in the instantaneous disruption of a large grain into small fragments. Since RATs are stronger for larger grains (\citealt{2007MNRAS.378..910L}; \citealt{2008MNRAS.388..117H}), RATD is more efficient for large grains than small ones. As shown in \cite{2019ApJ...876...13H}, the RATD mechanism is much faster than grain shattering and thus determines the upper cutoff of the grain size distribution in the ISM.

According to the RATD mechanism, the upper cutoff of grain size distribution is determined by the tensile strength, which depends on the grain internal structure (i.e., compact vs. composite structures; \citealt{2019ApJ...876...13H}). Unfortunately, the grain structure is one of the least constrained dust physical properties. In principle, one can constrain the internal structure with observational data if the variation of the polarization with the tensile strength is theoretically predicted \citep{2019ApJ...876...13H}. Therefore, the main goal of this paper is first to perform detailed modeling of multi-wavelength dust polarization from optical/UV to FIR/submm for the different local radiation intensity and grain tensile strength by simultaneously taking into account the alignment and rotational disruption of grains by RATs.

Full-sky polarization data from {\it Planck} have provided invaluable information about dust properties, grain alignment, and magnetic fields. A high polarization degree observed from the diffuse and translucent clouds by {\it Planck} reveals that dust grains must be efficiently (perfectly) aligned, which is consistent with a unified alignment theory of grains with iron inclusions \citep{2016ApJ...831..159H}. However, a detailed analysis of the polarization data for various clouds by \cite{2018arXiv180706212P} shows that the polarization degree at $\lambda=850\mum$ ($P_{850}$) does not always increase with grain temperature ($T_{d}$) as expected from the classical RAT theory. 
Instead, the polarization degree decreases with the grain temperature for $T_{d}\gtrsim 19\K$. This observed feature was thought to be a challenge to the classical picture of RAT alignment theory. However, as we will show in the paper, this unexpected trend would provide a valuable constraint on the tensile strength and then the internal structure of interstellar grains.

The structure of the paper is organized as follows. In Section \ref{sec:method}, we describe RAT alignment and RATD mechanism and our theoretical models to be used for modeling. In Sections \ref{sec:Pabs} and \ref{sec:Pem}, we will calculate the alignment size, disruption size for the different radiation fields, and present our modeling results of polarization of starlight and polarized emission. In Section \ref{sec:discussion} we will discuss the important implications of our study, focusing on the first constraint of grain internal structure with {\it Planck} data and the understanding of anomalous polarization of type Ia supernovae. A summary of our main findings are shown in Section \ref{sec:summary}.

% ========== Grain size distribution & Temperature ========== 
\section{Grain alignment and grain disruption by Radiative Torques}\label{sec:method}
% ---------- GRAIN ALIGNMENT
%\normalsize
In this section, we briefly review the theory of grain alignment and rotational disruption by RATs.

\subsection{Grain Alignment by RATs}\label{sec:Align}
\subsubsection{Critical size of aligned grains}
Let $u_{\lambda}$ be the spectral energy density of some radiation field. The total energy density is $u_{\rm rad}=\int u_{\lambda}d\lambda$. For the average interstellar radiation field (ISRF) in the solar neighborhood from \cite{1983AA...128..212}, one obtains the energy density $u_{\rm ISRF}=8.64\times 10^{-13}$erg cm$^{-3}$ and the mean wavelength $\bar{\lambda}=1.2\mum$ (\citealt{1997ApJ...480..633D}). Assuming that the radiation spectrum $u_{\lambda}$ is the same as the ISRF, one can describe the radiation energy density at a given location in the ISM by a dimensionless parameter $U=u_{\rad}/u_{\rm ISRF}$, which is referred to as {\it radiation strength}.

\begin{figure}
\centering
\includegraphics[scale=0.4]{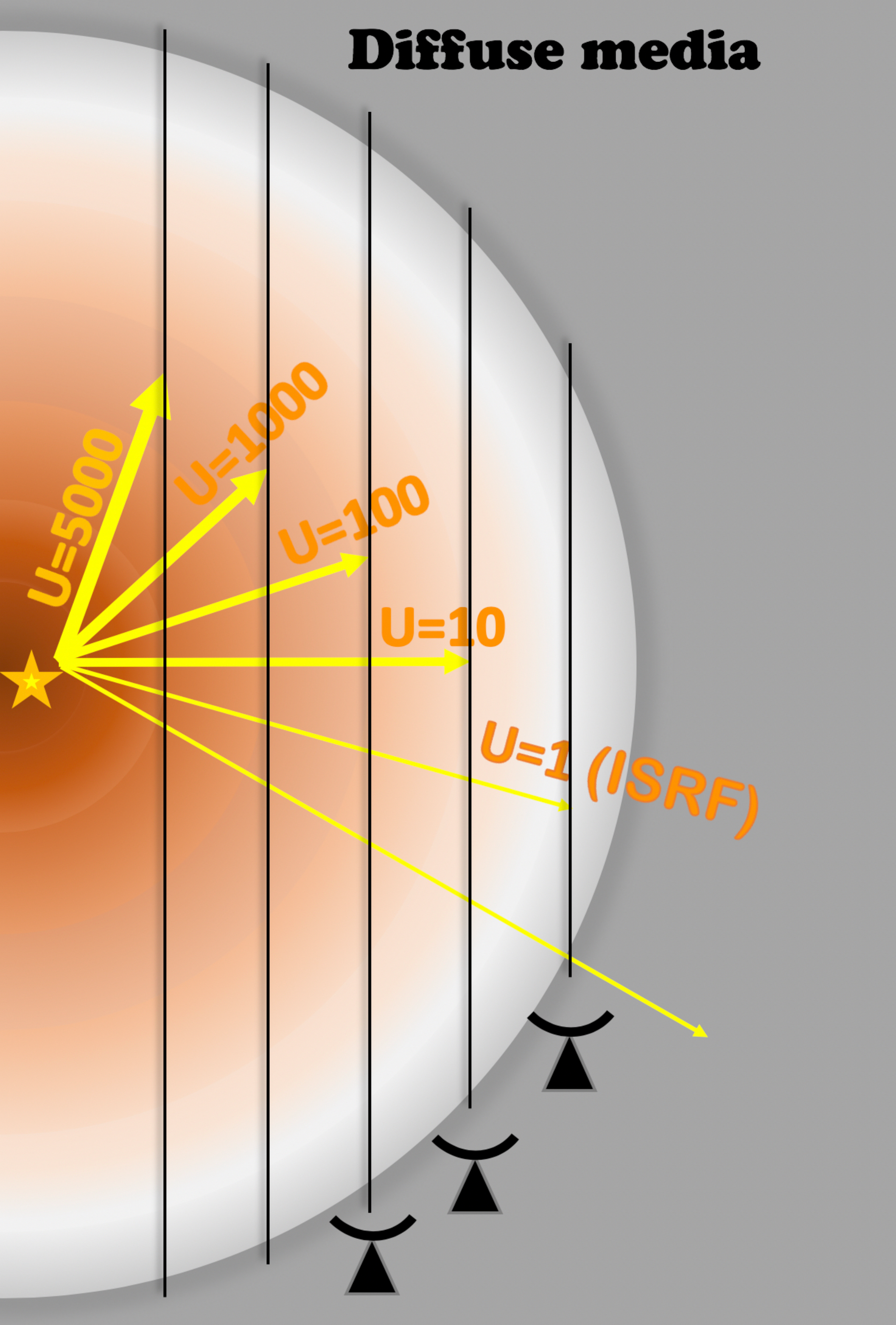}
\caption{Schematic illustration of a molecular cloud irradiated by a central star. Different lines of sights probe the different average radiation fields, which is characterized by the radiation strength $U$ spanning from $5000$ to $1$.}
\label{fig:GMC}
\end{figure}

To account for the variation of the local radiation intensity in the ISM, we will consider a wide range of the radiation strength for both the diffuse ISM and a molecular cloud illuminated by a nearby star as depicted in Figure \ref{fig:GMC}. We assume that a line of sight close to the star probes grains exposed to an averaged radiation field of strength $U=5000$, as illustrated in Figure \ref{fig:GMC}. Other lines of sight more distant from the star probe grains irradiated by weaker radiation fields. Note that the upper value of $U$ is chosen arbitrarily, but it is perhaps typical for photodissociation regions (PDRs).

Let $a$ be the effective size of the irregular grain which is defined as the radius of the equivalent sphere with the same volume as the irregular grain. According to the RAT alignment theory, grains are efficiently aligned when they can be spun-up to suprathermal rotation by an anisotropic radiation field. 

The radiative torque induced by the interaction of the anisotropic radiation field with the irregular grain is defined by
\begin{equation} % ----- "equation" of RATs
\Gamma_{\rm RAT} = \pi a^2\gamma u_{\rm rad}\left(\frac{\lambda}{2\pi}\right)Q_{\Gamma},
\label{eq:RATs}
\end{equation}
where $\gamma$ is the anisotropy degree of the radiation field, and $Q_{\Gamma}$ is the RAT efficiency. We adopt $\gamma=0.1$ for the diffuse medium and $\gamma=0.7$ for MCs (\citealt{1996ApJ...470..551D}).

Following \cite{2007MNRAS.378..910L}, the RAT efficiency can be approximately described by two power laws:

\begin{eqnarray}
Q_{\Gamma}\approx 0.4\left(\frac{\lambda}{1.8a}\right)^{-\eta},
\label{eq:Qgam}
\end{eqnarray}
where $\eta=0$ for $\lambda/a<1.8$ and $\eta=3$ for $\lambda/a\gtrsim 1.8$. This scaling was obtained by approximating numerical calculations with Equations (\ref{eq:Qgam}) for different grain compositions and grain shapes (\citealt{2007MNRAS.378..910L}). A slightly shallower slope is obtained from numerical calculations for an extended ensemble of grain shapes by \cite{2019ApJ...878...96H}.

The grain rotation is damped due to collisions with gas species (atoms and molecules) followed by their evaporation and the emission of IR photons after absorption of starlight. Let us define the ratio of the rotational gas damping to IR damping times as  $\tau_{\rm gas}/\tau_{\rm em}\equiv F_{\rm IR}$. By plugging $\Gamma_{\rm RAT}$ (Eq. \ref{eq:RATs}) into Equation (\ref{eq:omega_rad}), we can obtain the maximum angular velocity spun up by RATs: 

\begin{eqnarray}
\frac{\omega_{\rm RAT}}{\omega_{T}} &\simeq& 48.7\hat{\rho}a_{-5}^{3.2} U^{2.4}\left(\frac{\gamma}{0.1}\right)  \left(\frac{30\cm^{-3}}{n_{\rm H}}\right)\left(\frac{\overline{\lambda}}{1.2\mum}\right)^{-1.7}
\nonumber\\
&&\times \left(\frac{100\K}{T_{\rm gas}}\right) \left( 1+F_{\rm IR}\right),
\label{eq:wRAT}
\end{eqnarray} 
where $a_{-5}\equiv a/(10^{-5}\cm)$, $\hat{\rho}=\rho/(3\g\cm^{-3})$ with $\rho$ grain mass density, $F_{\rm IR}$ the dimensionless coefficient of rotational damping by IR emission as given by (see Appendix \ref{sec:damping})
\begin{equation}
F_{\rm IR} \simeq 0.4\left(\frac{U^{2/3}}{a_{-5}}\right)  
\left(\frac{30\rm{cm}^{-3}}{n_{\rm H}}\right)\left(\frac{100\K}{T_{\rm gas}}\right)^{1/2},
\end{equation}
and the thermal rotation rate $\omega_{\rm T}$ is given by
\begin{equation} 
\begin{split}
\omega_{\rm T}&= \left(\frac{kT_{\gas}}{I}\right)^{1/2}=\left(\frac{15kT_{\gas}}{8\pi\alpha_1\rho a^5}\right)^{1/2}\nonumber\\
&\simeq 1.6\times 10^{5}T_{2}^{1/2}a_{-5}^{-5/2} \alpha_{1}^{-1/2}\rm rad\s^{-1},
\end{split}
\label{eq:wT}
\end{equation}
where $T_{2}=T_{\gas}/100\K$, assuming the rotational kinetic energy of a grain around one axis is equal to $kT_{\gas}/2$. For simplicity, we assume $\alpha_{1}=1$ throughout the paper. For above analytical estimates, the RAT efficiency averaged over the radiation spectrum $\overline{Q}_{\Gamma}\approx 2(\bar{\lambda}/a)^{-2.7}$ for $a<\bar{\lambda}/1.8$ has been used (see Eq. 68 in \citealt{2014ApJ...790....6H}).

Let $a_{\rm align}$ be the critical size that grains can be driven to suprathermal rotation at which  $\omega_{\rm RAT}/\omega_{T}=3$. Above this limit, the degree of grain alignment starts to rise, and eventually grains achieve perfect alignment if high-J attractors are present (\citealt{2016ApJ...831..159H}). From Equation (\ref{eq:wRAT}) and the suprathermal rotation criterion, we can calculate the critical size of aligned grains for the various value of $U$. This alignment size depends on the gas density and temperature and the intensity in the radiation field. The representative results for a few values of $U$ with two different tensile strengths are listed in the Table \ref{tab:DustSize}. For the typical ISM, $U=1$, and one has $a_{\rm align} \sim 0.055\mum$, and $a_{\rm align}$ becomes smaller for higher $U$.
% alignment function

\begin{figure}
\includegraphics[scale=0.45]{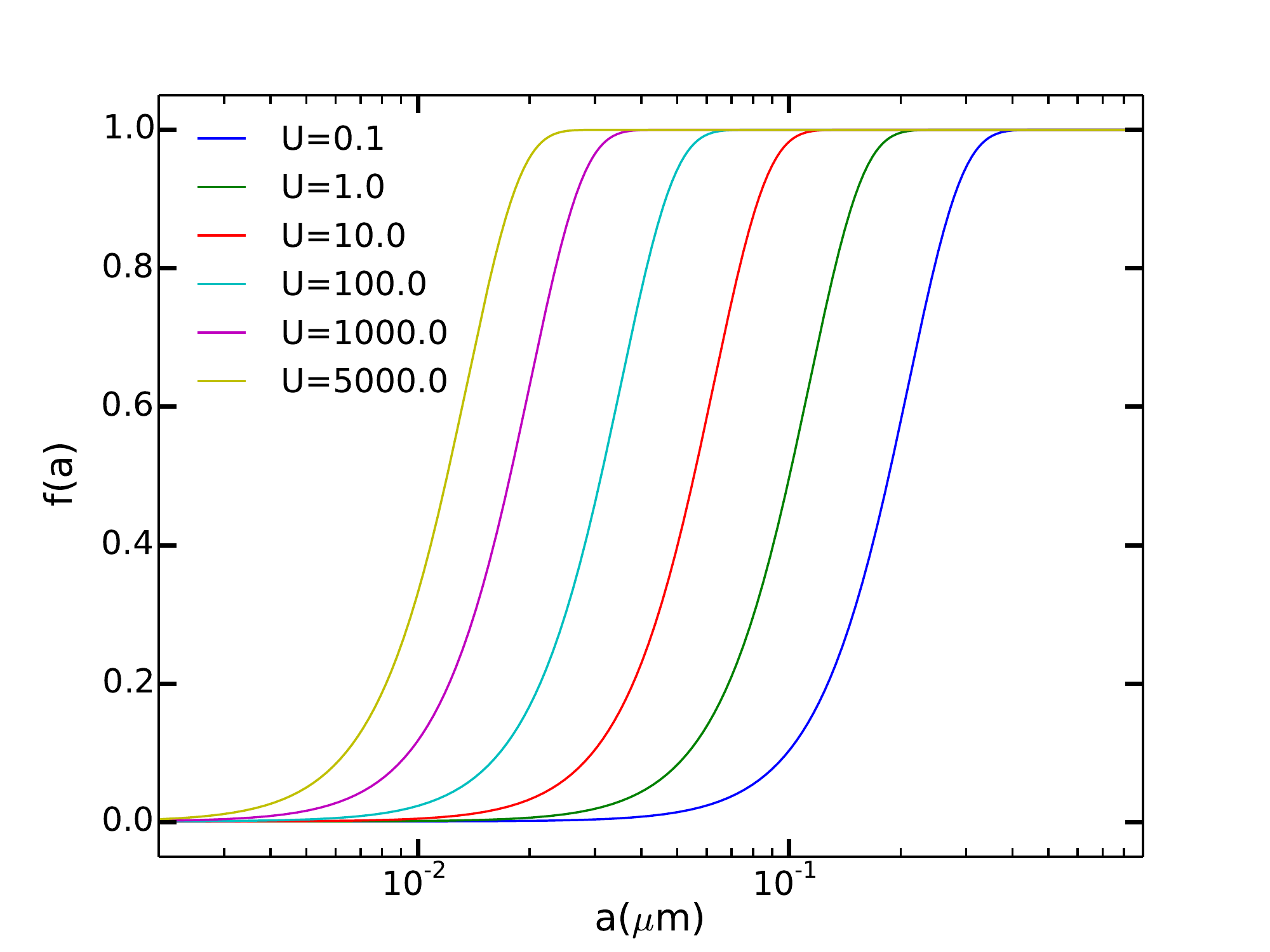}
\caption{Alignment function $f(a)$ obtained for various radiation strengths $U$. The alignment function is broader for higher $U$ due to the decrease of the alignment size $a_{\rm align}$ with increasing $U$.}
\label{fig:Afunc}
\end{figure}

\begin{table}[]
\centering
\begin{threeparttable} 
\caption{Physical parameters for the diffuse ISM and MC}
\label{tab:condition}
\begin{tabular}{l|cc}
\toprule
Parameters & Diffuse ISM & MC \\
\hline
n$_{\rm{\H}}$ (cm$^{-3}$)  & 30	& 10$^{4}$ \\
T$_{\rm{\gas}}$ (K)				   & 100				& 20  \\
$\rho$ (g/cm$^{3}$)			    &  3				 &  3  \\
$\gamma$					    &  0.1				 &  0.7  \\
$u_{\rm rad}$ (\erg\cm$^{-3}$) 	   &  8.64 x 10$^{-13}$     &  Varied  \\
$\bar{\lambda}$ ($\mu$m)			   &  1.2			 & Varied  \\
\hline
\end{tabular}
\end{threeparttable}
\end{table}

\subsubsection{Grain alignment function}
Numerical simulations in \cite{2016ApJ...831..159H} show that if the RAT alignment has a high-J attractor point, then, grains can be perfectly aligned when they are spun-up to suprathermal rotation. For grains without iron inclusions (i.e., ordinary paramagnetic material), high-J attractors are only present for a limited range of the radiation direction that depends on the grain shape. This range is increased if grains have an enhanced magnetic susceptibility by including iron clusters (\citealt{2008ApJ...676L..25L}; \citealt{2016ApJ...831..159H}).
For grains smaller than $a_{\rm align}$, numerical simulations show that the alignment degree is rather small even in the presence of iron inclusions because grains rotate subthermally \citep{2016ApJ...831..159H}. Thus, to describe the size-dependence of grain alignment degree, we adopt an alignment function
\begin{equation}
f(a)=f_{\rm min} + (f_{\rm max}-f_{\rm min})[1-\exp(-(\frac{0.5a}{a_{\rm align}})^{3}],
\end{equation}
where we fix $f_{\rm max} = 1.0$ corresponding to the perfect alignment of large grains. The lower value $f_{\rm min}$ describes the alignment of small grains of $a<a_{\rm align}$, chosen to be $10^{-3}$, so that their contribution to the total polarization is negligible.

The above alignment function approximately agrees with the results obtained from inverse modeling of polarization data from \cite{2009ApJ...696....1D} and \cite{2014ApJ...790....6H}. Therefore, it is appropriate to use this alignment function for modeling dust polarization.

Figure \ref{fig:Afunc} shows the alignment function calculated for the different radiation strength. One can see that the stronger radiation field can align smaller grains, shifting the alignment function toward smaller sizes. In the other word, the range of aligned grains becomes broader for higher radiation intensity.

%\footnote{$f=0$ when grains are randomly aligned. $f=1$ when the the short axis of the grain is perfectly aligned along a magnetic field.}

\begin{table}[]
\begin{center}
\caption{Grain alignment and disruption size for the diffuse media}
\begin{tabular}{c|c|c|c|c}
\toprule
\multirow{4}{*}{Radiation Strength} & \multicolumn{4}{c}{Diffuse ISM} \\
\cline{2-5}
 & \multicolumn{2}{c|}{$S_{\max}$=10$^7$ erg cm$^{-3}$} & \multicolumn{2}{c}{$S_{\max}$=10$^9$ erg cm$^{-3}$} \\
 \cline{2-5}
 & \textbf{a}$\boldsymbol{_{\mathrm{align}}}$ & \textbf{a}$\boldsymbol{_{\mathrm{disr}}}$ & \textbf{a}$\boldsymbol{_{\mathrm{align}}}$ & \textbf{a}$\boldsymbol{_{\mathrm{disr}}}$ \\
(U) & ($\mu$m) & ($\mu$m) & ($\mu$m) & ($\mu$m) \\
\hline
0.1      & 0.105 & 1.0     & 0.105 & 1.0\\
1         & 0.057 & 0.31   & 0.057 & 1.0\\
10       & 0.031 & 0.15   & 0.031 & 0.4\\
100     & 0.018 & 0.10   & 0.018 & 0.25\\
1000   & 0.010 & 0.076 & 0.010 & 0.18\\
5000   & 0.007 & 0.062 & 0.007 & 0.15      
\end{tabular}
\end{center}
%\begin{tablenotes}
%\item Notes. Grain size(\textbf{a}) : \textbf{a}$_{\mathrm{align}} \leq \hspace{0.1cm} $\textbf{a}$ \hspace{0.1cm} \leq \hspace{0.1cm} $\textbf{a}$_{\mathrm{disr}}$ 
%\end{tablenotes}
\label{tab:DustSize}
\end{table}

% ---------- Grain disruption size

\normalsize
\subsection{Grain rotational disruption by the RATD mechanism} \label{sec:Disrupt}
\subsubsection{Grain disruption size and tensile strength}
A rapidly spinning grain of angular velocity $\omega$ develops a tensile stress of $S=\rho \omega^{2}a^{2}/4$ with $\rho$ being the mass density of dust. For large grains in the strong radiation field, the angular velocity by RATs can be sufficiently large such that $S$ exceeds the maximum tensile strength $S_{\max}$ of grain material, resulting in rotational disruption (\citealt{2019NatAs...3..766H}). 

The critical angular velocity that the grain is disrupted is given by $S=S_{\max}$, which yields
\begin{equation} 
\omega_{\rm disr} \simeq \frac{3.6\times 10^8}{a_{-5}}S^{1/2}_{\max,7}\hat{\rho}^{-1/2} \hspace{0.3cm} \rm{rad\,s^{-1}},
\label{eq:wdisr}
\end{equation}
where $S_{\max,7}=S_{\max}/(10^{7}\erg\cm^{-3})$. 

The tensile strength measures the maximal mechanical limit that can resist against an applied tension force before it breaks. The exact value of $S_{\rm max}$ depends on the grain internal structure and composition, which is never constrained for interstellar dust. Physically, compact grains are expected to have higher $S_{\rm max}$ than composite grains due to the difference in the bonding energy between grain constituents. Thus, a higher $S_{\max}$ implies a more compact grain, while a lower value of $S_{\max}$ implies a porous or composite grain (\citealt{2019ApJ...876...13H}). For instance, polycrystalline bulk solid can have $S_{max} \sim 10^{9} - 10^{10} \erg \cm^{-3}$ (\citealt{Burke74}; \citealt{Draine79}), while ideal materials, i.e., diamond, have $S_{\max} \sim 10^{11} \erg \cm^{-3}$ (see \citealt{2019NatAs...3..766H}). In this paper, we assume a reasonable range of the tensile strength, $S_{\max}\sim 10^6 - 10^9$ erg cm$^{-3}$ to account for various structures of interstellar grains. 
By equating Equations (\ref{eq:wRAT}) and (\ref{eq:wdisr}), one can obtain the critical size $a_{\rm disr}$ above which grains are disrupted as follows:
\begin{equation}
\left( \frac{a_{\rm disr}}{0.1 \mu\rm{m}}\right)^{2.7}\simeq 5.1\gamma^{-1}_{-1}U^{-1/3}\bar{\lambda}^{1.7}_{0.5}S^{1/2}_{\max,7},
\label{eq:adisr}
\end{equation}
for strong radiation fields of $U\gg 1$ and $a_{\max}\le \bar{\lambda}/1.8$, and $\gamma_{-1}=\gamma/0.1$.

The maximum size that grains are still disrupted by RATD is given by (see \citealt{2019ApJ...877...36H})
\begin{eqnarray}
a_{\rm disr,max}&\simeq& 2.8\gamma\bar{\lambda}_{0.5}\left(\frac{U}{\hat{n}\hat{T}_{\gas}^{1/2}}\right)^{1/2}\left(\frac{1}{1+F_{\rm IR}}\right)\nonumber\\
&&\times \hat{\rho}S_{\max,7}^{-1/2}~\mum.\label{eq:adisr_up}
\end{eqnarray}

Equation (\ref{eq:adisr_up}) gives $a_{\rm disr,max}\sim 2.4\mum$ for the tensile strength of $S_{\max}\approx 10^{7}\erg\cm^{-3}$ and $\gamma_{\rm rad}=0.5$. Therefore, for the diffuse ISM, all grains between $a_{\rm disr}$ and $a_{\max}=1\mum$ are disrupted. Here we disregard the possibility of having micron-sized grains in the ISM, therefore, essentially all grains larger than $a_{\rm disr}$ are destroyed by RATD.

%  the molecular cloud with starless core and dust grains affected only by interstellar radiation field. Deeper, the effect of radiation on dust grains alignment is diminishing. We cannot see the disruption of dust grains by radiation and larger grains are aligned grain with larger visual extinction, AV.

Using numerical calculations, we can obtain the critical size of grain alignment ($a_{\rm align}$) and rotational disruption ($a_{\rm disr}$) by RATs for the various radiation field strength and local gas properties. Table \ref{tab:DustSize} lists the values of $a_{\rm align}$ and $a_{\rm disr}$ for dust grains in the diffuse ISM illuminated by the different radiation fields. In the typical ISM ($U=1$), dust grains of size $a\gtrsim 0.06\mum$ can be aligned by, whereas grains of $a\gtrsim 0.31\mum$ are disrupted by RATs. In Table \ref{tab:DustSize}, we see that both the alignment and the disruption size become smaller as the radiation field strength increases.

\subsubsection{Grain size distribution in the presence of RATD}\label{sec:GrainSIZE_T}

We adopt a mixed dust model consisting of two separate populations of amorphous silicate grains and carbonaceous (graphite) grains (see \citealt{2001ApJ...548..296W}; \citealt{2007ApJ...657..810D}. The grain size distribution of component $j=sil$ or $gra$ follows a power-law distribution (\citealt{1977ApJ...217..425M}, hereafter MRN):
\begin{equation}
\frac{1}{n_{\H}}\frac{dn_{j}}{da} = C_{j}a^{-3.5} \hspace{0.7cm} \rm{at} \hspace{0.1cm} \it a_{\rm min}<a<a_{\rm max},
\label{eq:MRN}
\end{equation}
where $dn_{\rm j}$ is the number density of grains of material $j$ between $a, a+da$, $n_{\rm H}$ is the number density of hydrogen, and $a_{\rm min}$=10{\AA} and $a_{\max}=a_{\rm disr}$ are assumed. We take constant $C_j$ from \cite{2001ApJ...548..296W} for MRN size distribution as follows: $C_{\rm sil}=10^{-25.11}$cm$^{2.5}$ and $C_{\rm gra}=10^{-25.13}$cm$^{2.5}$.

The RATD mechanism tends to reduce the abundance of large grains and increases the abundance of smaller grains because of the conservation of total dust mass. To account for this effect, we assume that the slope of the size distribution can be constant and increase the normalization constant $C$ (see more details in \cite{2019arXiv190611498C}).

Previous studies (e.g., \citealt{1995ApJ...444..293K}; \citealt{2006ApJ...652.1318D}; \citealt{2009ApJ...696....1D}) show that different grain shapes and axial ratio $r$ can reproduce the observational data. Therefore, we consider two special cases of a prolate spheroidal shape with $r=1/3$ and an oblate spheroidal shape with $r=1.5$, for both silicate and carbonaceous grains. 

% ========== Polarization by Absorption ========== 
 
\section{Polarization of Starlight}\label{sec:Pabs}

\subsection{Polarization Curves} 
For modeling polarization, we assume that graphite grains are randomly oriented, whereas silicate grains can be aligned via RATs. The polarization of starlight arising from absorption and scattering of aligned silicate grains in a slab of thickness $dz$ is defined as
\begin{eqnarray}
dp_{\lambda}(x,z)=\frac{1}{2} \int^{a_{\max}}_{a_{\rm align}} \left(C_x-C_y\right) \frac{dn_{sil}}{da}dadz,
\label{eq:dplam_xz}
\end{eqnarray}
where $C_{x}$ and $C_{y}$ are the grain cross section along the $x-$ and $y-$ axes, respectively, in the reference system which the line of sight is directed along the z-axis. 

Following \cite{2014ApJ...790....6H}, one has
\begin{eqnarray}
C_x - C_y = C_{\rm pol}R{\rm cos}^2\zeta,
\label{eq:cross-section}
\end{eqnarray}
where $R$ is the Rayleigh reduction factor (\citealt{1999MNRAS.305..615R}), and $\zeta$ is the angle between the magnetic field and the plane of the sky. 

Let $f=R{\rm cos}^2\gamma$ be the effective degree of grain alignment, which depends on the grain size (\citealt{2016ApJ...831..159H}). In the following, to explore how the polarization spectrum changes with the local radiation field, we assume that the magnetic field is uniform along the line of sight and lies in the plane of the sky, so $\cos^{2}\gamma=1$. The polarization degree produced by all grains along the line of sight is given by
\begin{eqnarray}
\frac{P_{\lambda}}{N_{\H}}=\int^{a_{\max}}_{a_{\rm align}} \frac{1}{2}C_{\pol}^{sil}(\lambda,a)f(a)\frac{1}{n_{\H}}\frac{dn_{sil}}{da}da,
\label{eq:Plam}
\end{eqnarray}
where $N_{\H}=n_{\H}L$ with $L$ the length of the line of sight. 

Equation (\ref{eq:Plam}) can be rewritten as
\begin{equation} 
P({\lambda})=\sigma_{\pol}(\lambda)\times N_{\H},
\label{eq:sigpol}
\end{equation}
where $\sigma_{\pol}$ in units of $\cm^{2} \H^{-1}$ is the polarization cross section.

\subsection{Numerical Results}
To get insights into the dependence of the polarization spectrum on grain alignment and disruption, we consider two typical environments, the standard diffuse ISM and dense molecular clouds with the physical parameters listed in Table \ref{tab:condition}. We calculate the polarization of starlight using the cross-section $C_{\rm pol}$ and $C_{\rm ext}$ obtained from \cite{2018A&A...610A..16G} which fit well the average Planck full-sky emission and polarization.

\begin{figure*}
\includegraphics[scale=0.45]{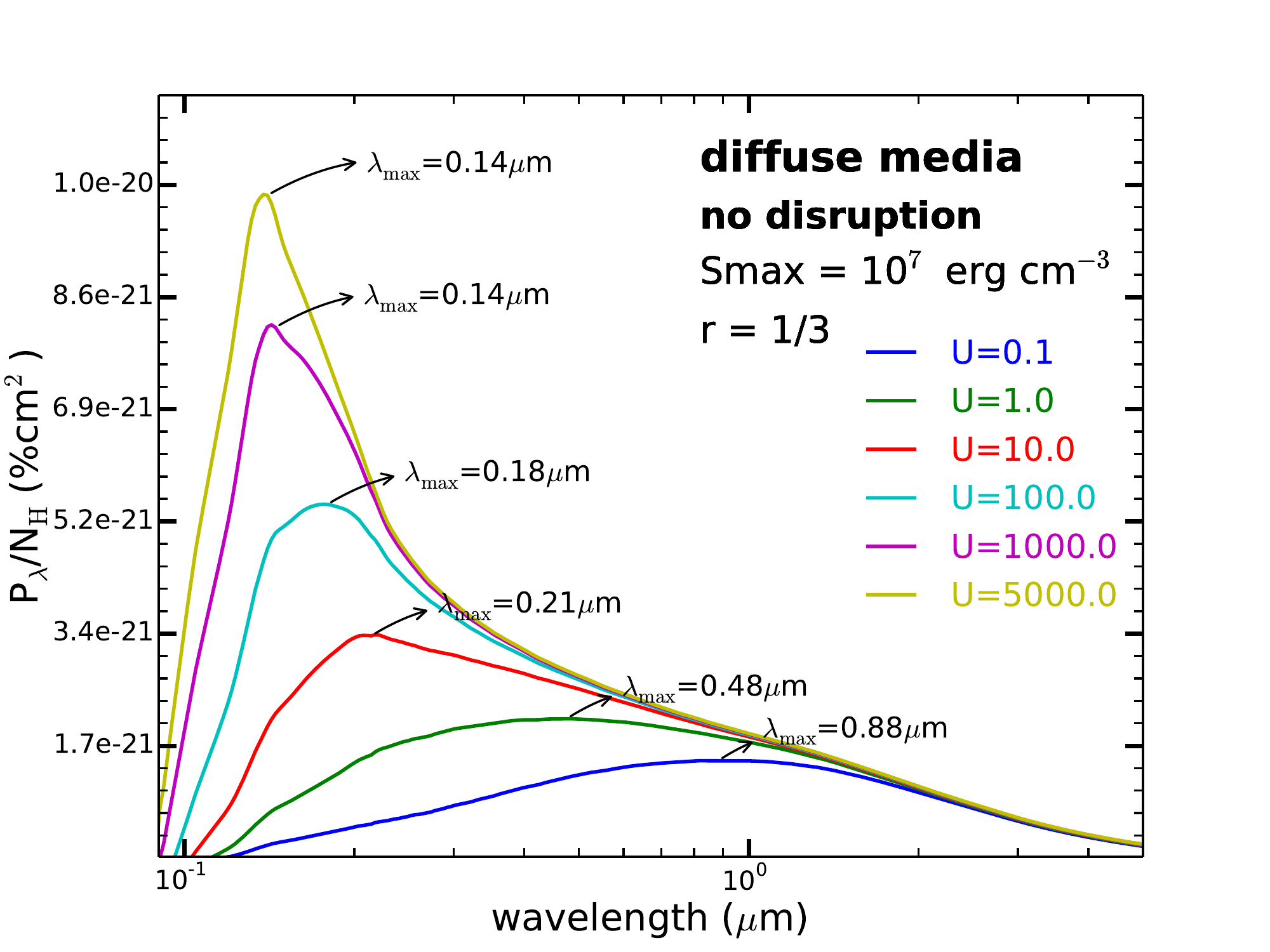}
\includegraphics[scale=0.45]{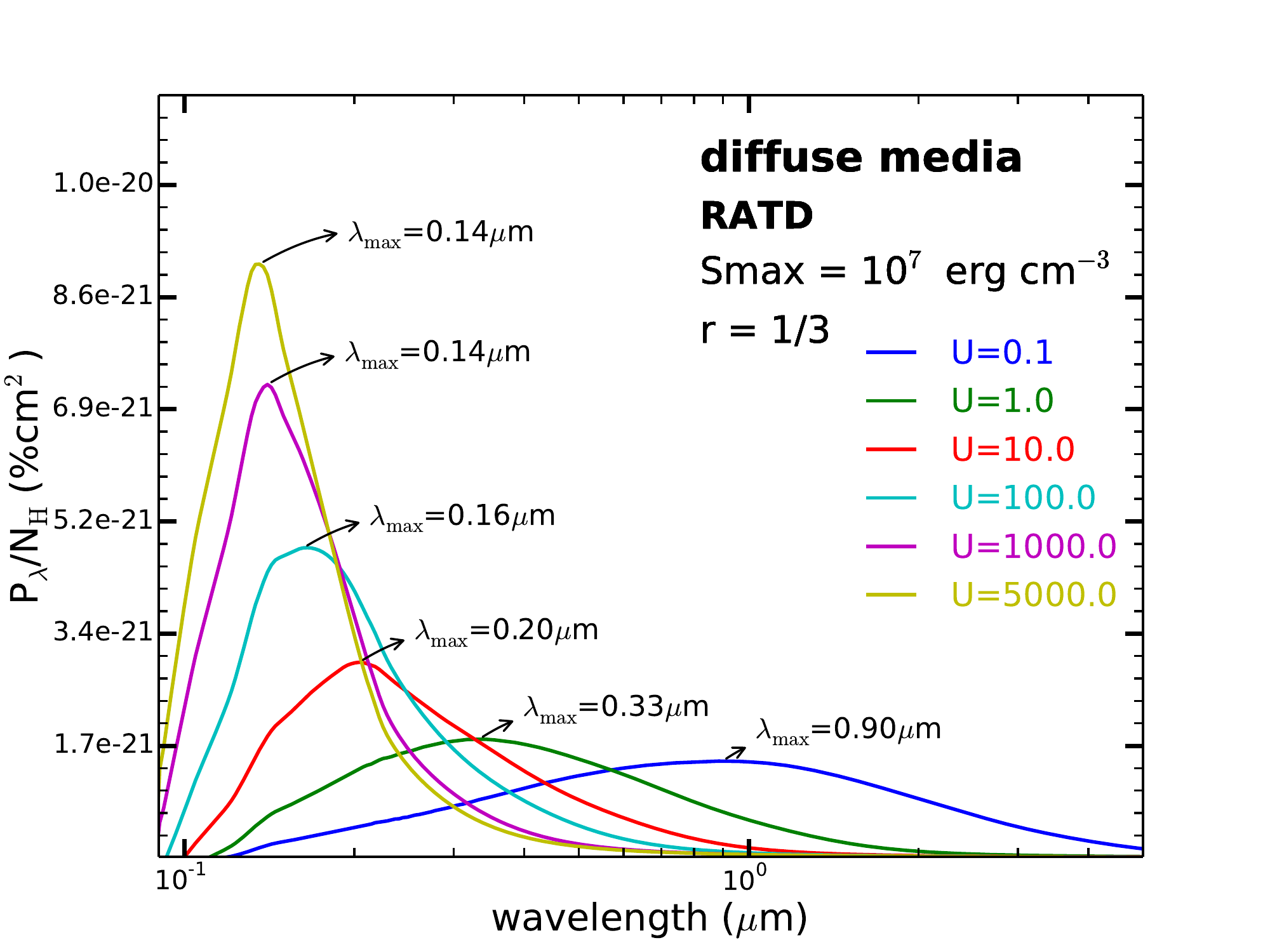}
\caption{Polarization spectrum due to extinction of starlight by dust grains aligned with axial ratio $r=1/3$ by RATs for the diffuse media with various radiation field strengths for two cases without RATD and with RATD. The tensile strength $S_{\max}=10^{7}\erg\cm^{-3}$ is considered.}
\label{fig:PabsDiff}
\end{figure*}

Figure \ref{fig:PabsDiff} shows the polarization curves for the diffuse ISM without RATD (left panels) and with RATD (right panel), assuming an axial ratio of grains $r=1/3$ and the tensile strength $S_{\max}=10^{7}\erg\cm^{-3}$.

Figure \ref{fig:PabsDiff} (left panel) shows that the polarization spectrum with $r=1/3$ in the diffuse media peaks at $\lambda_{\max} \sim 0.48 \mu$m when RATD is not taken into account. The polarization at $U=1$ reflects the polarization spectrum from the typical interstellar radiation field. As the radiation field strength increases, $\lambda_{\max}$ moves to shorter wavelengths because of the enhancement of small grains. 

Figure \ref{fig:PabsDiff} (right panel) shows the results obtained when RATD is taken into account. It shows that the width of the polarization spectrum becomes narrower as the radiation field strength increases. The reason for that is that the stronger radiation field not only makes smaller grains to be aligned but also disrupts large grains into smaller ones. As a result, the polarization at long wavelength (optical-NIR) decreases, and the polarization at UV wavelengths increases with increasing $U$.

\begin{figure*}
\includegraphics[scale=0.45]{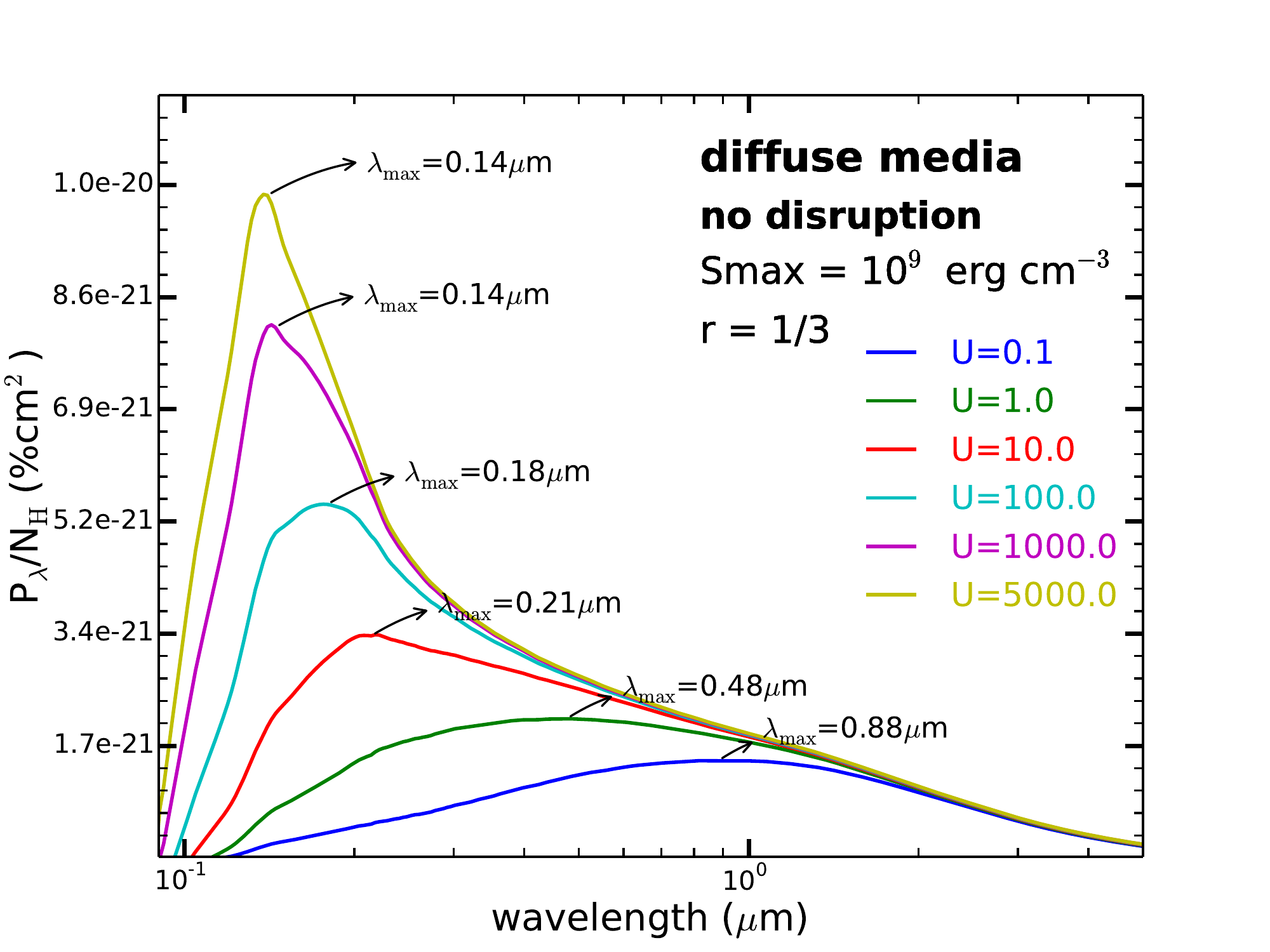}
\includegraphics[scale=0.45]{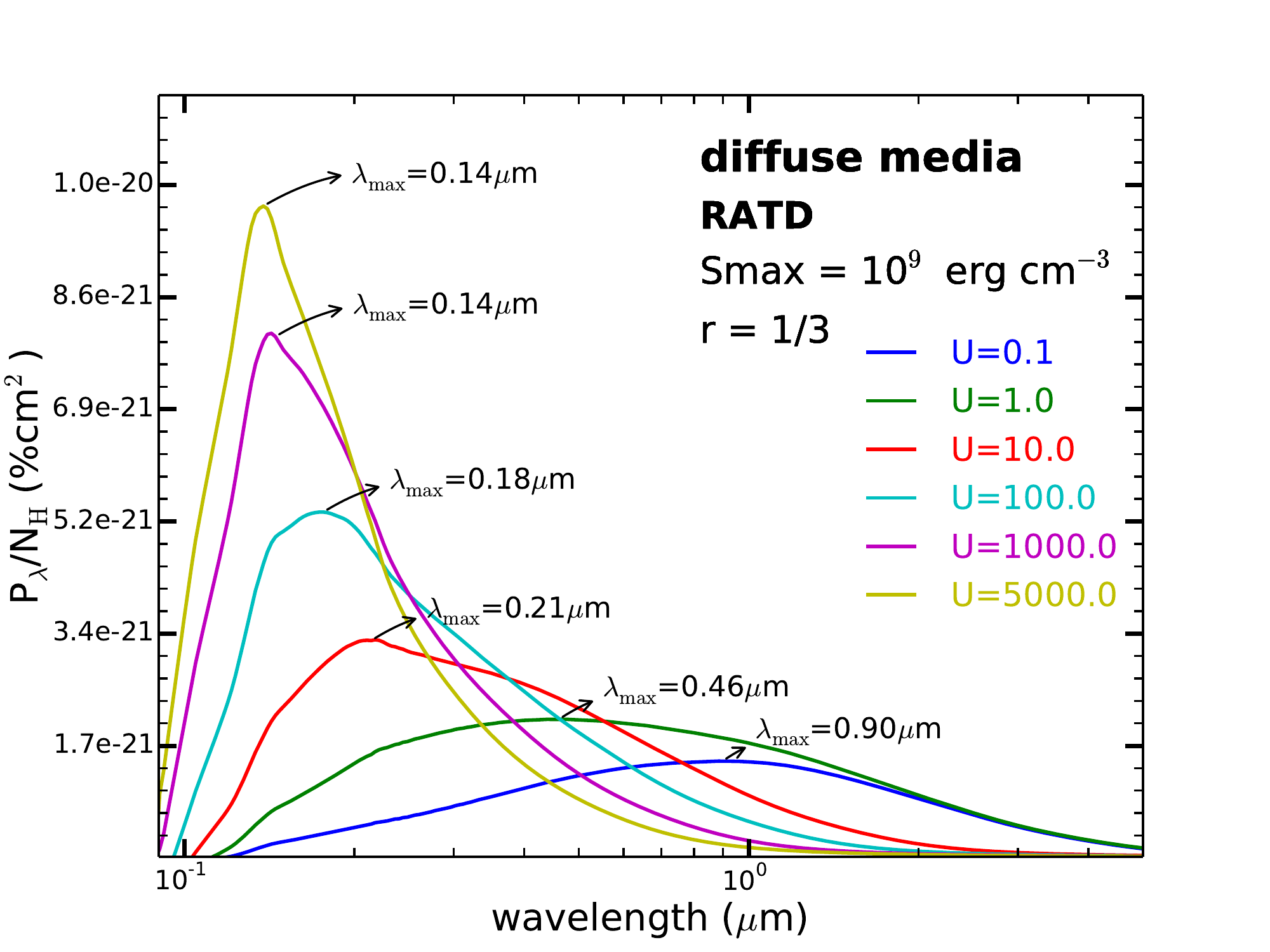}
\caption{Same as Figure \ref{fig:PabsDiff} but for $S_{\max}=10^{9}\erg\cm^{-3}$}
\label{fig:Pabsdiff_Smax2}
\end{figure*}

Same as Figure \ref{fig:PabsDiff}, but Figure \ref{fig:Pabsdiff_Smax2} shows the results for a higher tensile strength. The results for the case without disruption is the same, but the effect of RATD (right panel) is less prominent than the case of the lower tensile strength.

\begin{figure*}
\includegraphics[scale=0.45]{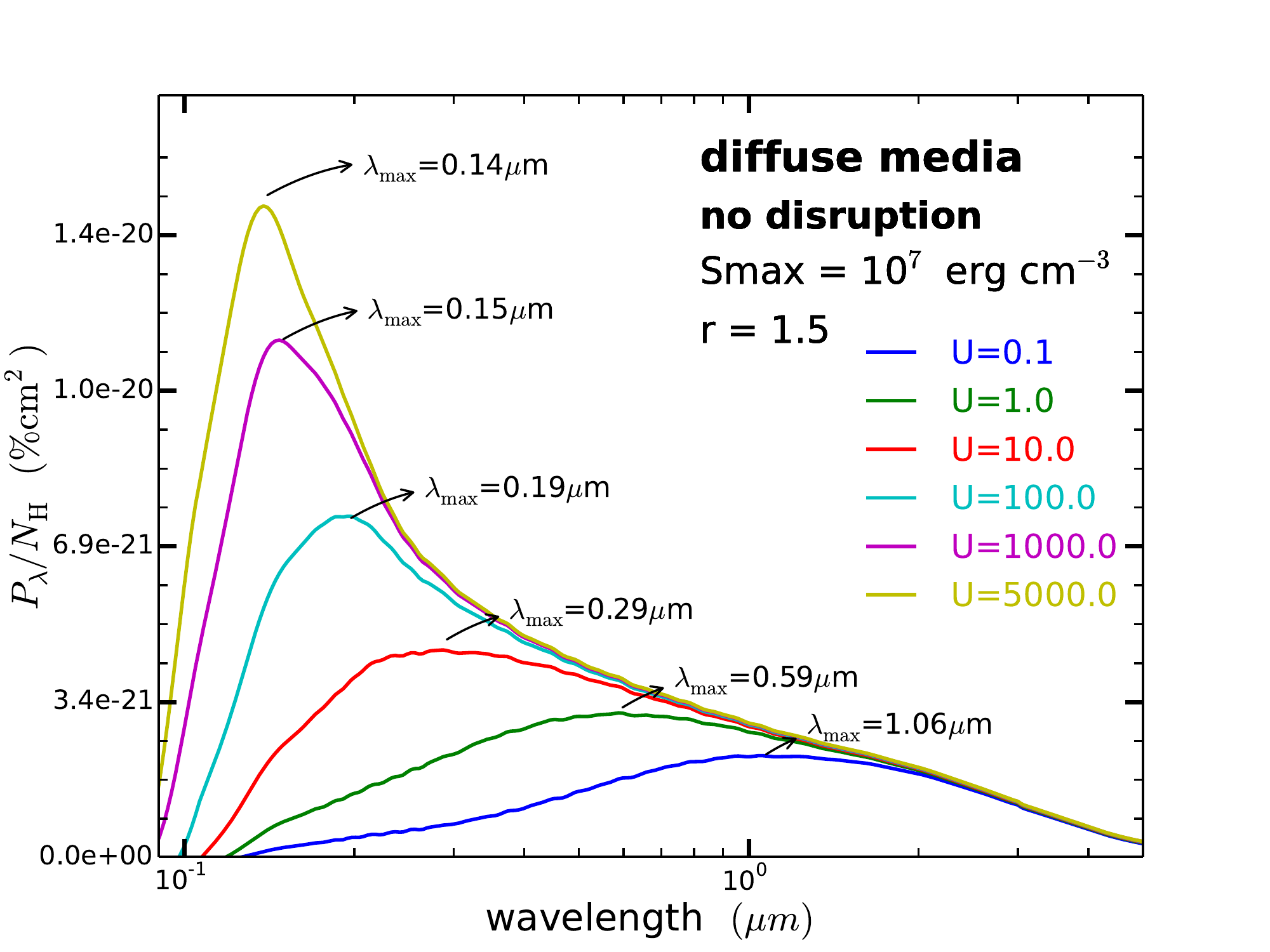}
\includegraphics[scale=0.45]{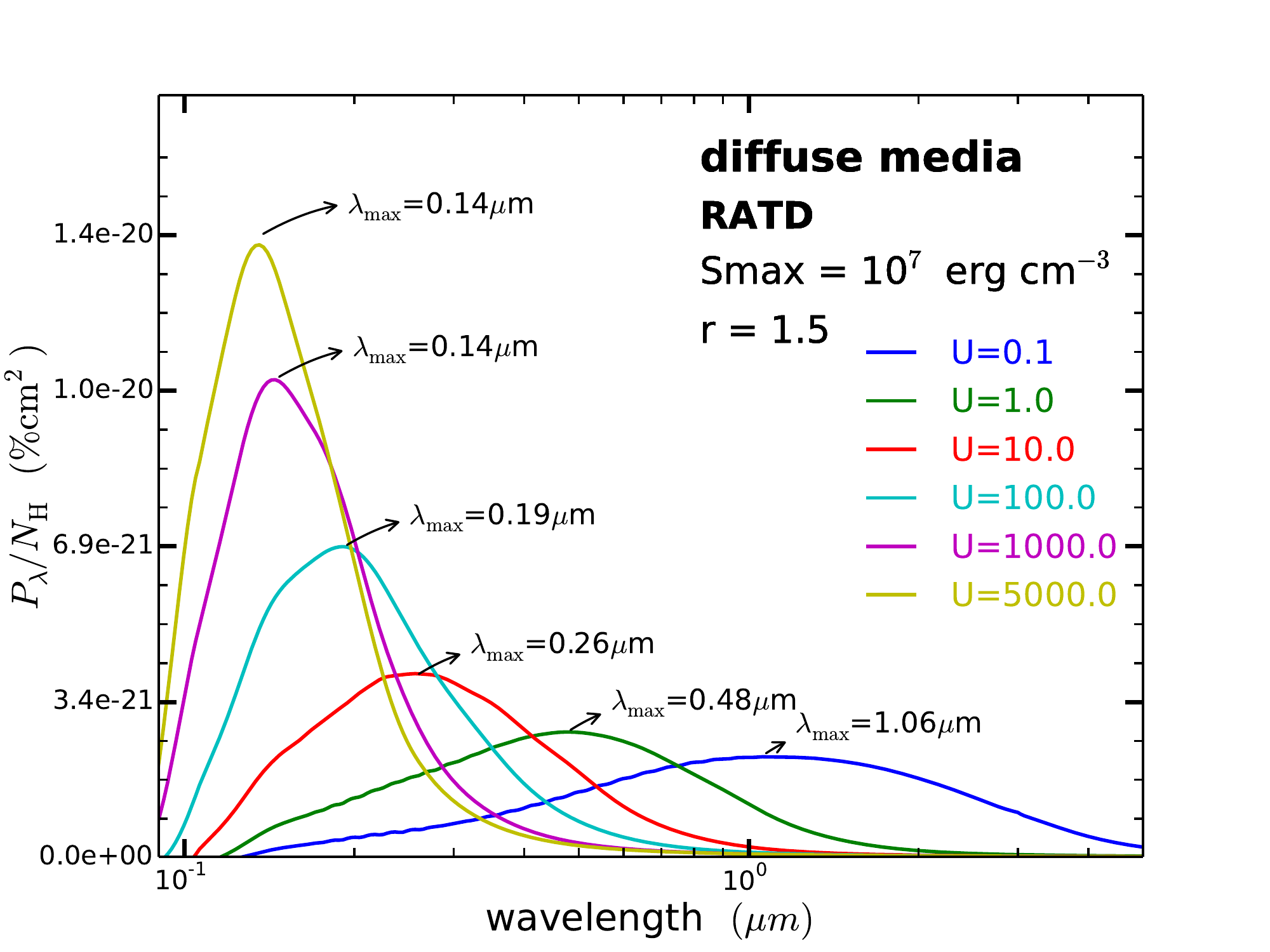}
\caption{Same as Figure \ref{fig:PabsDiff} but for oblate grains with axial ratio $r=1.5$.}
\label{fig:PabsDiffr15}
\end{figure*}

Figure \ref{fig:PabsDiffr15} shows similar results but for an axial ratio $r=1.5$. The maximum polarization is larger than for the case $r=1/3$, but the peak wavelength is similar.

Following Equation (\ref{eq:adisr}), the grain disruption size by RATD is determined by the tensile strength. The radiation disrupts larger grains when they have a higher tensile strength and the wide range of aligned grain size distribution contribute to polarization. The bottom and right panels of Figure \ref{fig:PabsDiff} show the polarization from grains with $S_{\max}$=10$^9$ $\erg\cm^{-3}$ that the width of the polarization spectrum is larger than that for $S_{\max}$=10$^7$ $\erg\cm^{-3}$. On the other hand, the polarization curves have a similar shape regardless of the tensile strength when the RATD mechanism is not applied, as shown in the bottom and left panels of Figure \ref{fig:PabsDiff}.

\begin{figure*}
\includegraphics[scale=0.45]{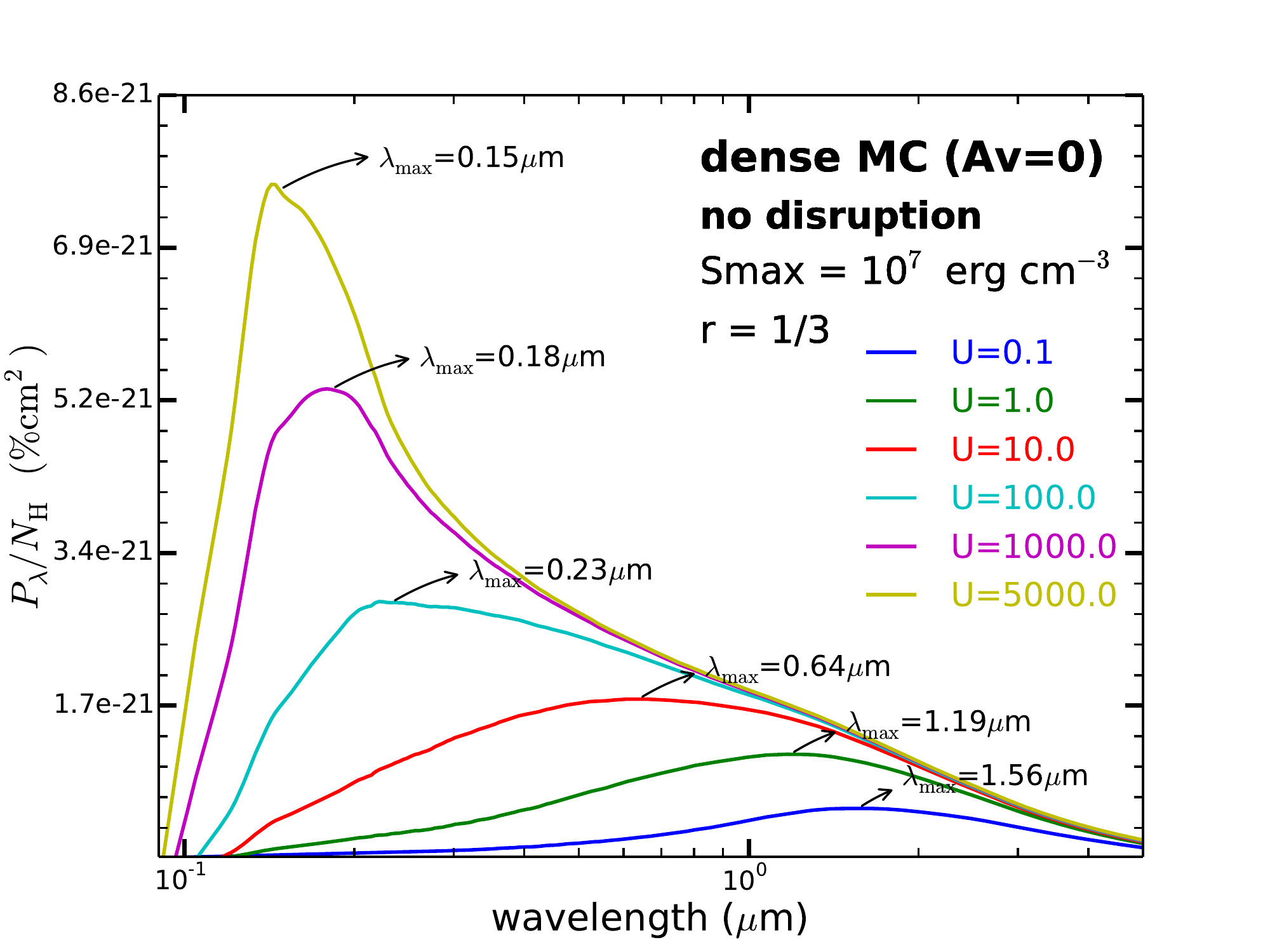}
\includegraphics[scale=0.45]{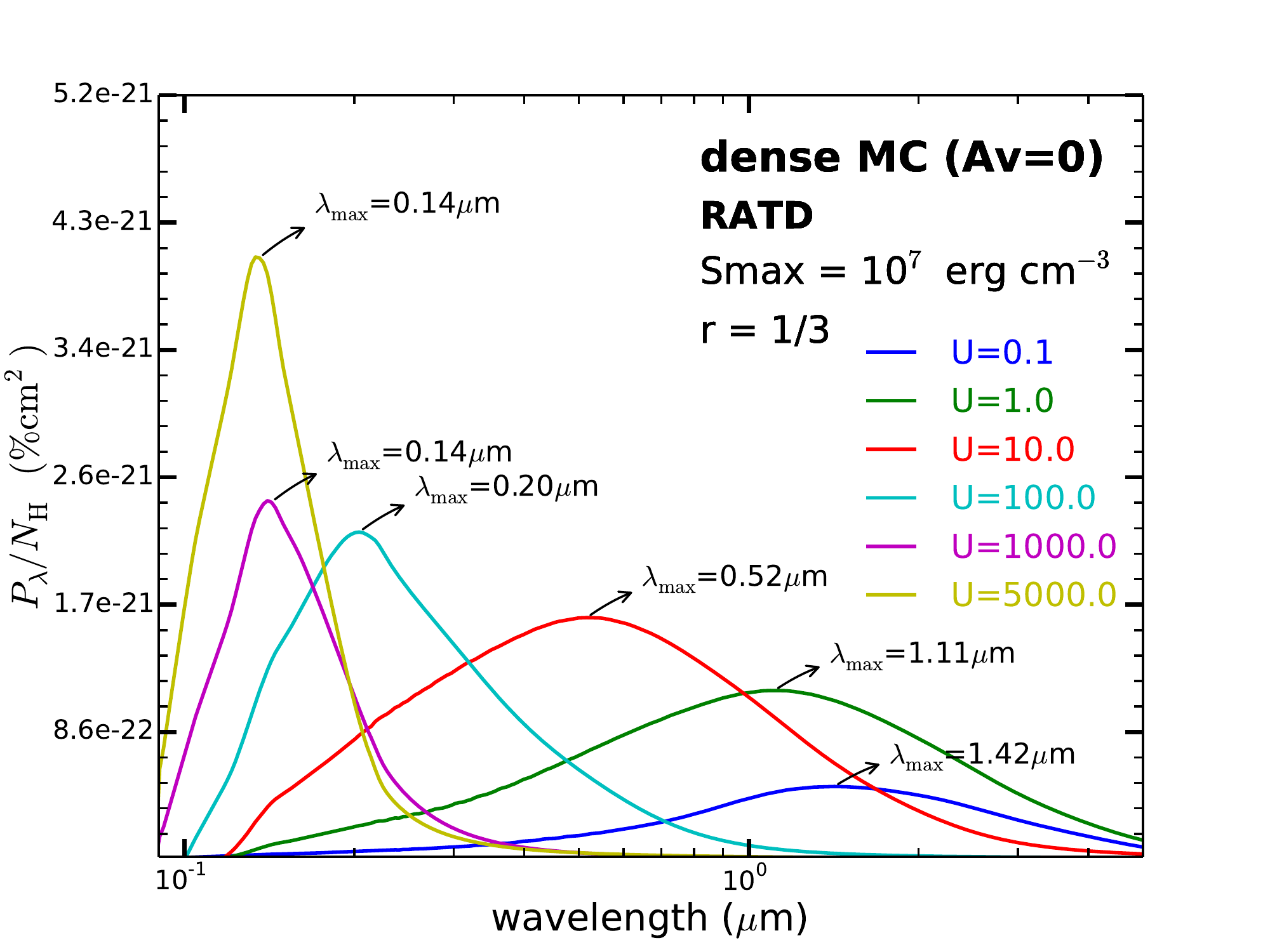}
\caption{Polarization spectrum due to extinction of starlight by aligned grains in a molecular cloud by dust grains with axial ratio of $r=1/3$ and the tensile strength of $10^{7}\erg\cm^{-3}$ for the case without RATD (left panel) and with RATD (right panel). The polarization spectrum changes with different $U$.}
\label{fig:PabsGMC}
\end{figure*}

\begin{figure*}
\includegraphics[scale=0.45]{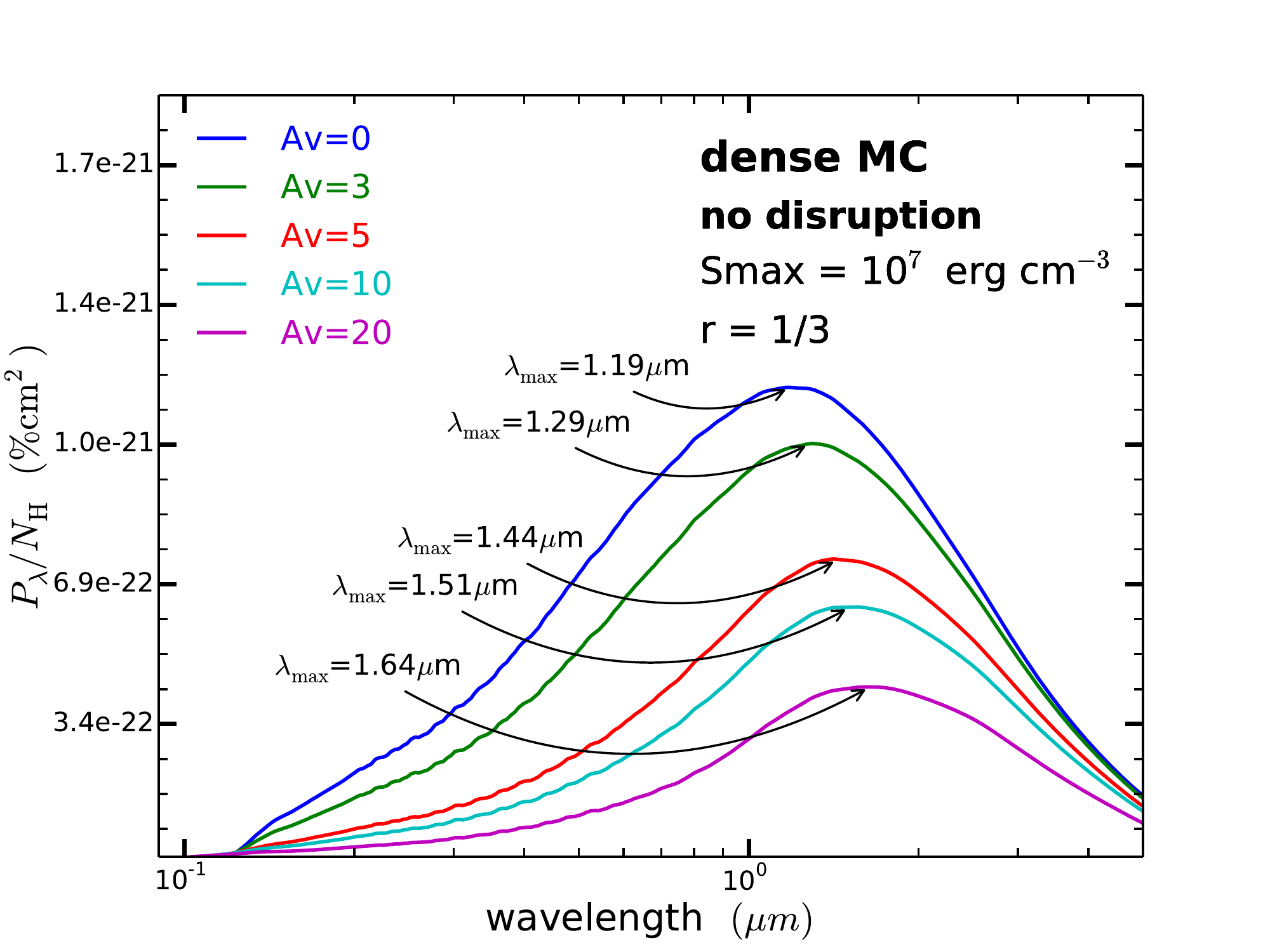}
\includegraphics[scale=0.45]{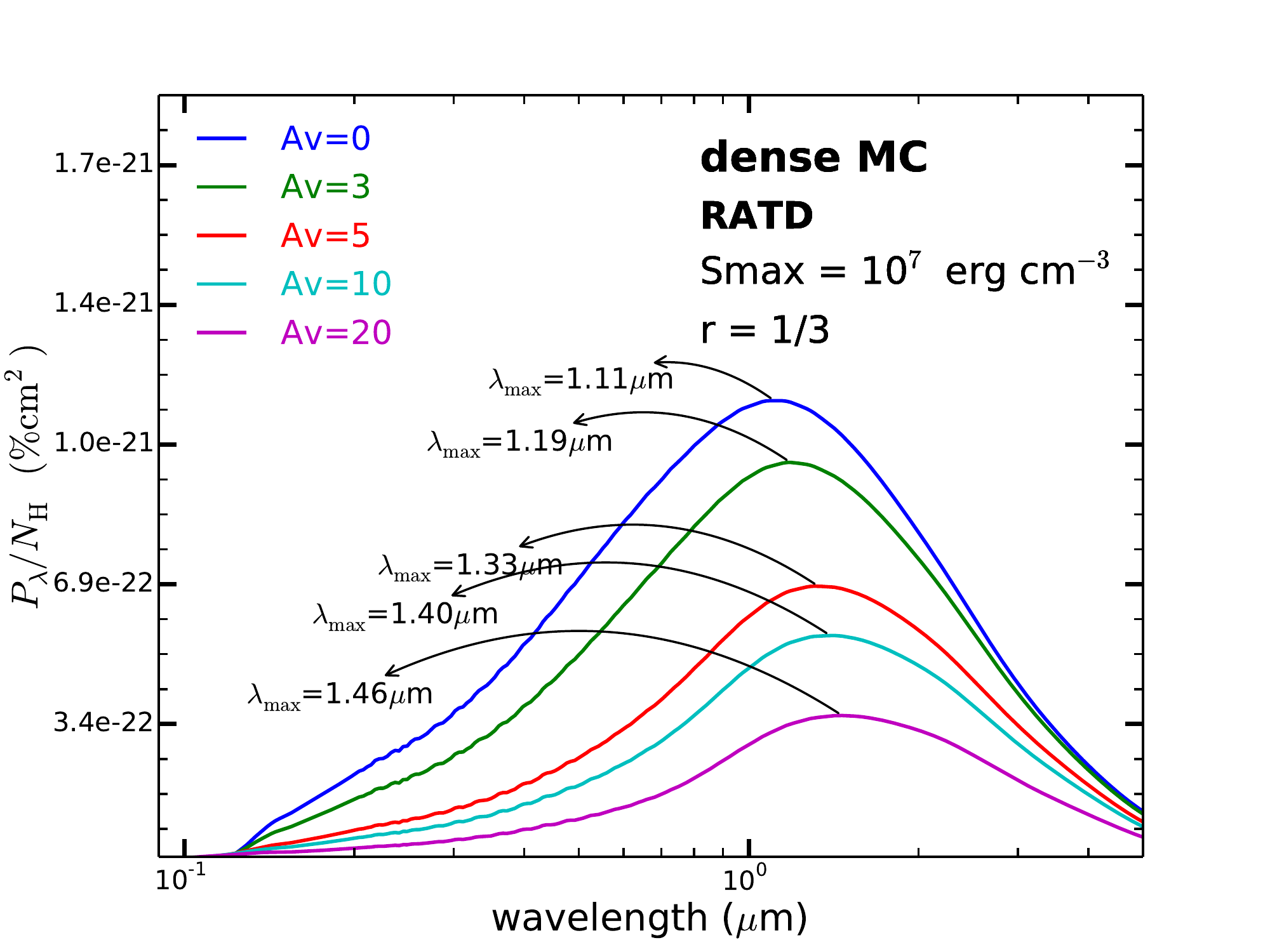}
\includegraphics[scale=0.45]{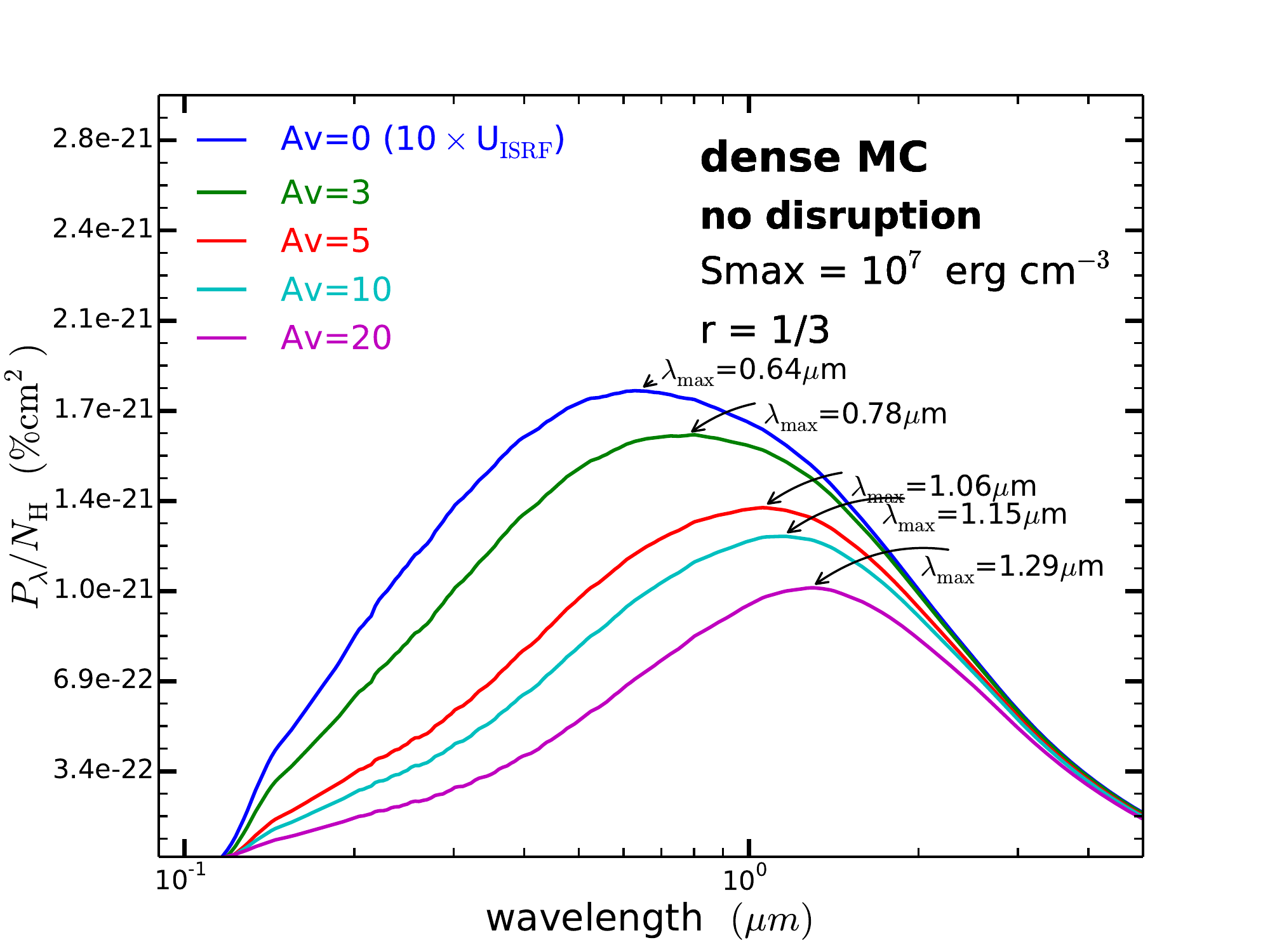}
\includegraphics[scale=0.45]{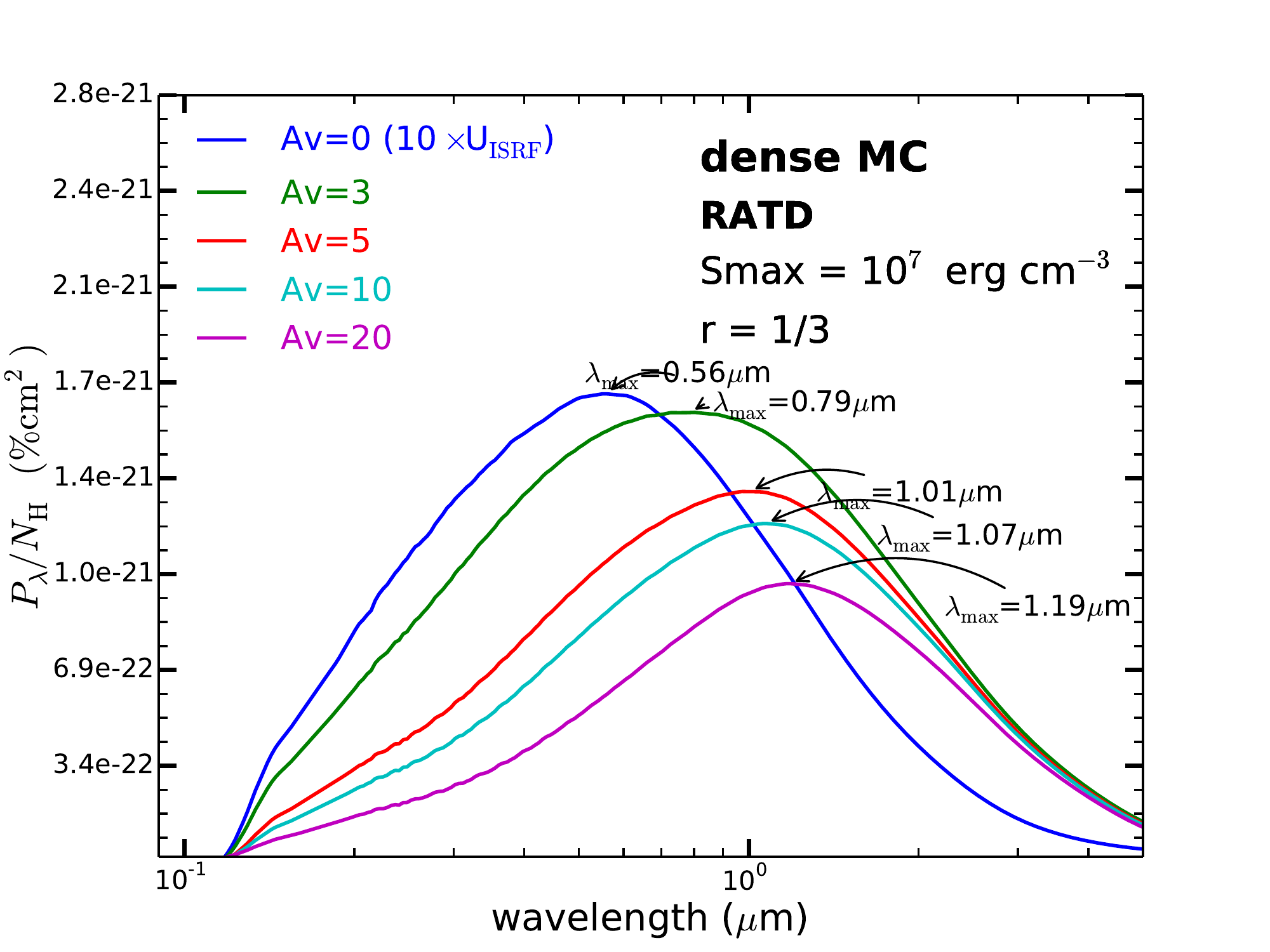}
\caption{Same as Figure \ref{fig:PabsGMC} but for oblate grains with axial ratio $r=1.5$. Two different strength of interstellar radiation field at $A_V$=0 is used: the strength of radiation at $A_V$=0 is a typical ISM radiation field (1$\times u_{\rm ISRF}$) in top panels and it is 10$\times u_{\rm ISRF}$ in bottom panels.}
\label{fig:PabsGMC_Av}
\end{figure*}

Figure \ref{fig:PabsGMC} and \ref{fig:PabsGMC_Av} show the results for a molecular cloud with a star and without a star at the center, respectively. The polarization fraction is lower than that in the diffuse media (Figure \ref{fig:PabsDiff}) because the high number gas density results in a faster rotational damping time such that the critical size of aligned grains is larger. As a result, the maximum polarization is decreased, and the peak wavelength is increased. One can see from the top panels of Figure \ref{fig:PabsGMC_Av} that the polarization curves has little change of the peak wavelength as the visual extinction increases and almost same at high $A_V$, where very weak radiation results in the disruption of grains very little. 

We also consider the case where the ambient interstellar radiation field is 10 times stronger than the standard ISRF, such that grains at $A_V$=0 in a dense MC are exposed to $U=10$. The obtained results are shown in the bottom panels of Figure \ref{fig:PabsGMC_Av}. The profile of the polarization spectrum is similar to the top panels, but the polarization at longer wavelengths at $A_V$=0 becomes smaller when RATD is taken into account (see right panels of Figure \ref{fig:PabsGMC_Av}). It arises from the fact that large grains at the surface of the molecular cloud can be disrupted into small grains only by a stronger radiation field. However, deep inside the cloud, even the strong interstellar radiation cannot disrupt large grains due to dust extinction.

% ========== Polarization by Emission ========== 

\section{Polarized Thermal Emission from Dust Grains}\label{sec:Pem}
\subsection{Polarization degree}
Dust grains heated by starlight re-emit thermal radiation in infrared. For the optically thin regime, the total emission intensity and polarized intensity are respectively given by (\citealt{2009ApJ...696....1D}):
\begin{equation} 
\begin{split}
\frac{I_{\rm em} (\lambda)}{N_{\H}} &= \sum_{j=sil,car}\int ^{a_{\max}}_{a_{\rm min}}Q_{\rm ext} \pi a^2 \int dT B_{\lambda}(T_d)\frac{dP}{dT} \frac{1}{n_{\H}}\frac{dn_{j}}{da} da,\\
\frac{I_{\rm pol} (\lambda)}{N_{\H}} &= \int ^{a_{\max}}_{a_{\rm min}}f(a)Q_{\rm pol}\pi a^2 \int dT B_{\lambda}(T_d)\frac{dP}{dT} \frac{1}{n_{\H}}\frac{dn_{sil}}{da} da,
\end{split}
\label{eq:Ipol_Iem}
\end{equation}
$dP/dT$ is the temperature distribution function which depends on the grain size and radiation strength $U$, and $B_{\lambda}$ is the Planck function as given by
\begin{equation} 
B_{\lambda}(\lambda, T) = \frac{2hc^2}{\lambda^5}\frac{1}{e^{hc/(k_B T\lambda)}-1}.
\label{eq:blackbody}
\end{equation}
Above, we disregard the minor effect of grain alignment on the thermal emission, which is considered in \cite{2009ApJ...696....1D}.

The polarization degree is then given by
\begin{equation} 
P (\%)= 100\times \left(\frac{I_{\rm pol}}{I_{\rm em}}\right).
\label{eq:Pem_ratio}
\end{equation}

%\begin{figure*}
%\includegraphics[scale=0.45]{results/IemDiffd.pdf}
%\includegraphics[scale=0.45]{results/IemGMCd.pdf}
%\caption{Total emission intensity by dust grains with various %radiation field strengths in diffuse media (left), or in dense %media (right).}
%\label{fig:Iem}
%\end{figure*}
% ---------- Grain Temperature
\subsection{Grain temperature distribution} \label{sec:GrainT}

\begin{figure*} 
\includegraphics[scale=0.45]{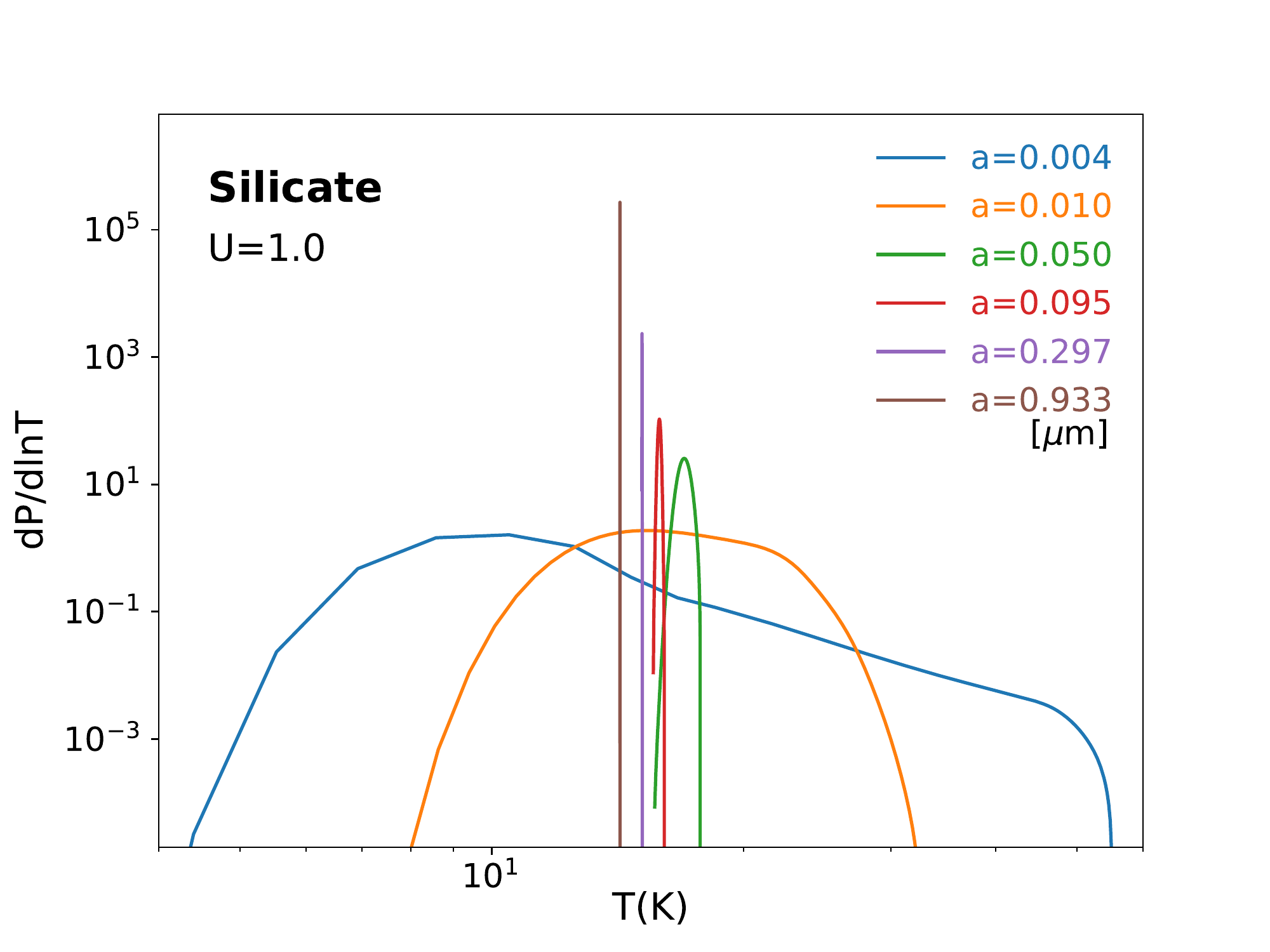}
\includegraphics[scale=0.45]{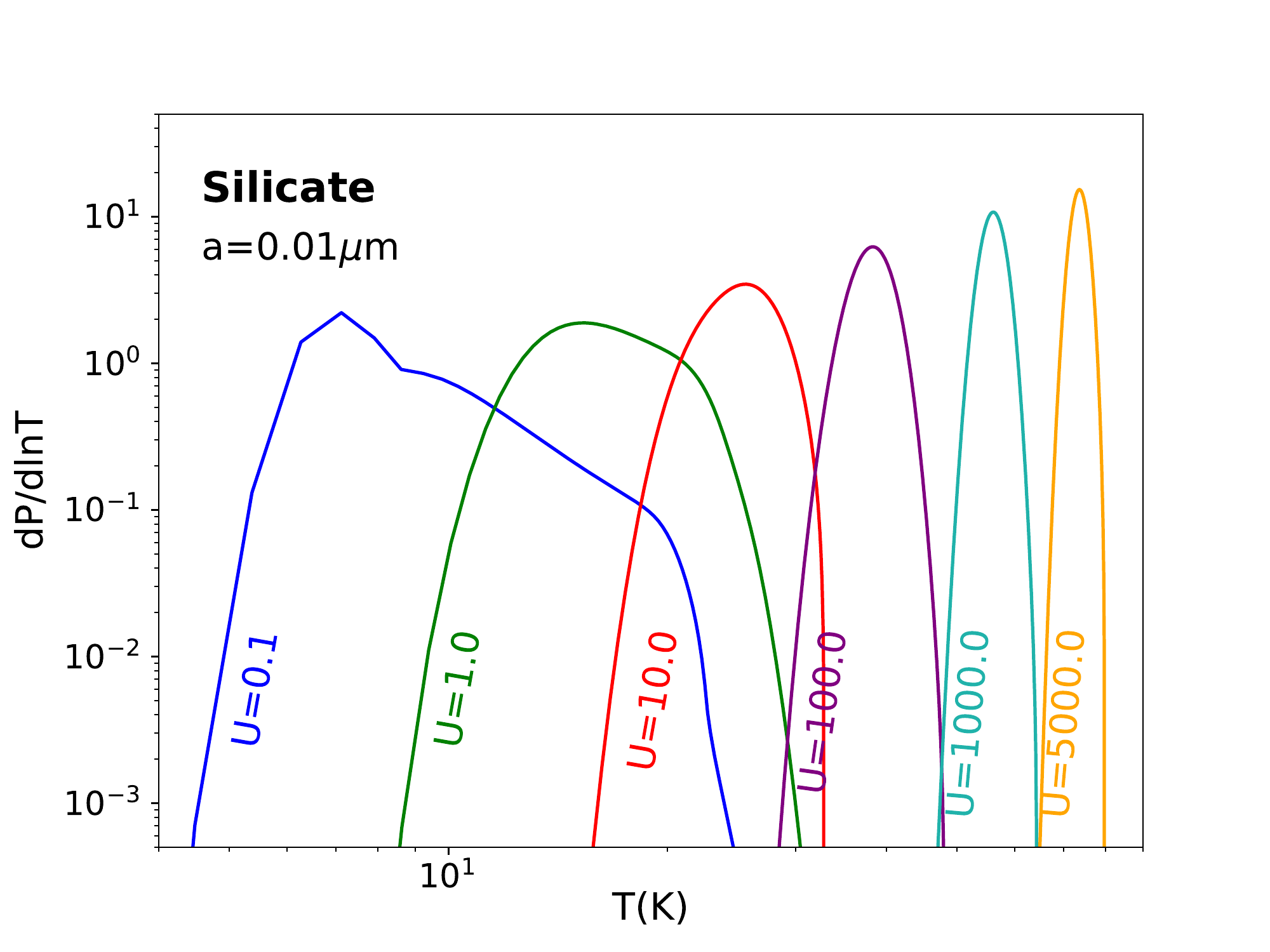}
\caption{Temperature probability distribution $dP/d\rm{ln}T$ for silicate grains at $U=1$ with various grain sizes (\it{left panel}) and at $a\sim 0.01\mu$m with $U = 0.1 - 5000$ (\it{right panel}).}
\label{fig:Tdust}
\end{figure*}

Dust grains are heated to high temperatures by absorption of optical/UV photons from stars, and subsequently, the grains cool down by re-emitting photons at long wavelengths. Let $dP$ be the probability of finding the grain temperature in the interval $[T, T+dT]$. Large grains can achieve a steady temperature due to high heat capacity, but small grains undergo strong temperature fluctuations due to its low heat capacity.

We compute $dP/dT$ using a DustEM code which is publicly available at https://www.ias.u-psud.fr/DUSTEM/. Figure \ref{fig:Tdust} (left panel) shows the temperature distribution function of silicate grains at several sizes and the standard radiation field ($U=1$). The temperature distribution is very broad for small grains ($a<0.05\mum$) and becomes narrower for larger grains. The right panel of Figure \ref{fig:Tdust} shows the change in the temperature for silicate grains of size $a=0.01\mum$ subject to various radiation fields. For a low radiation strength of $U<10$, the temperature distribution is broad, and the distribution becomes narrower and shifts to higher peak temperature as $U$ increases.

\subsection{Polarization spectrum for the diffuse interstellar medium}
\begin{figure*}
\includegraphics[scale=0.45]{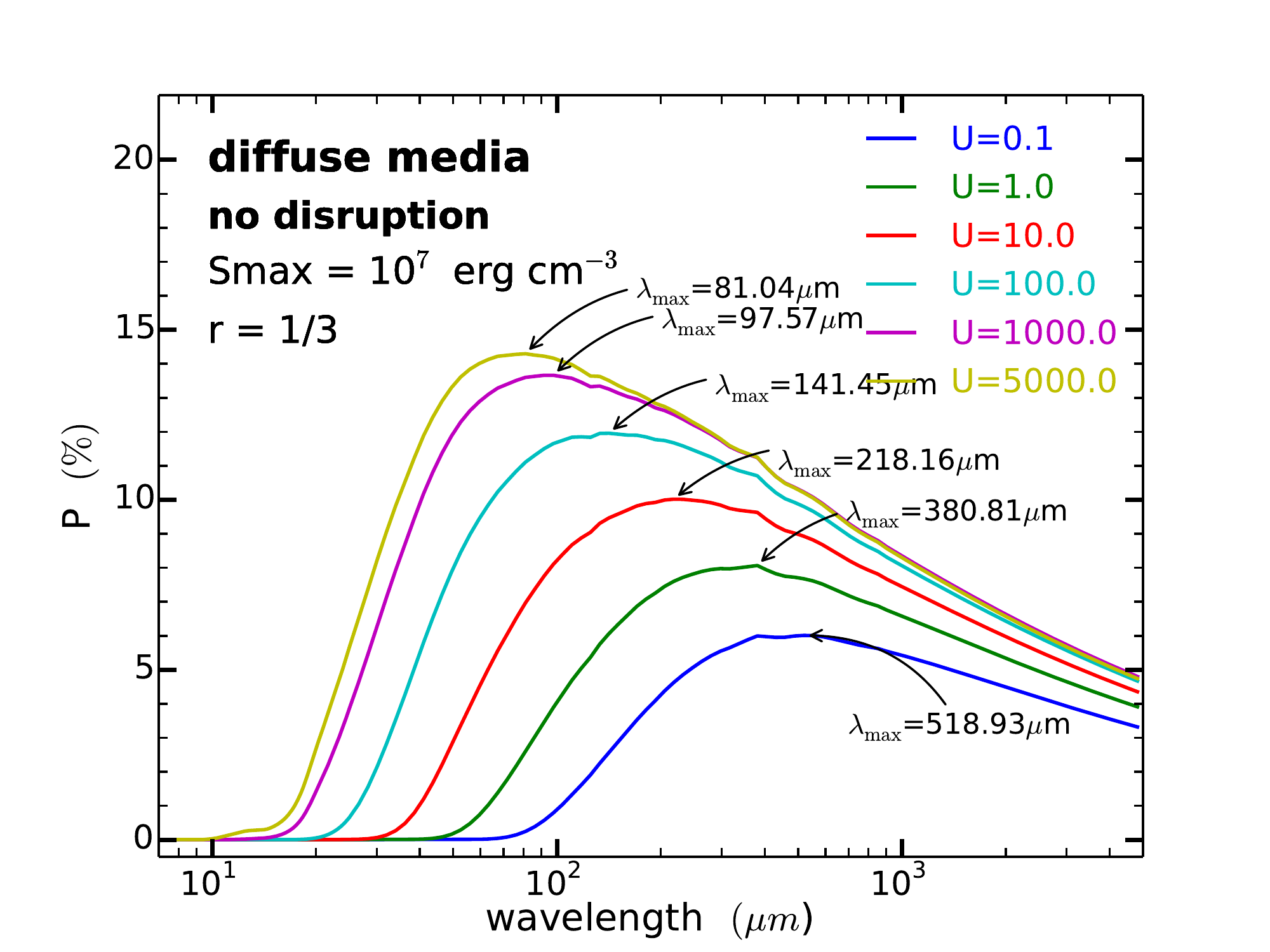}
\includegraphics[scale=0.45]{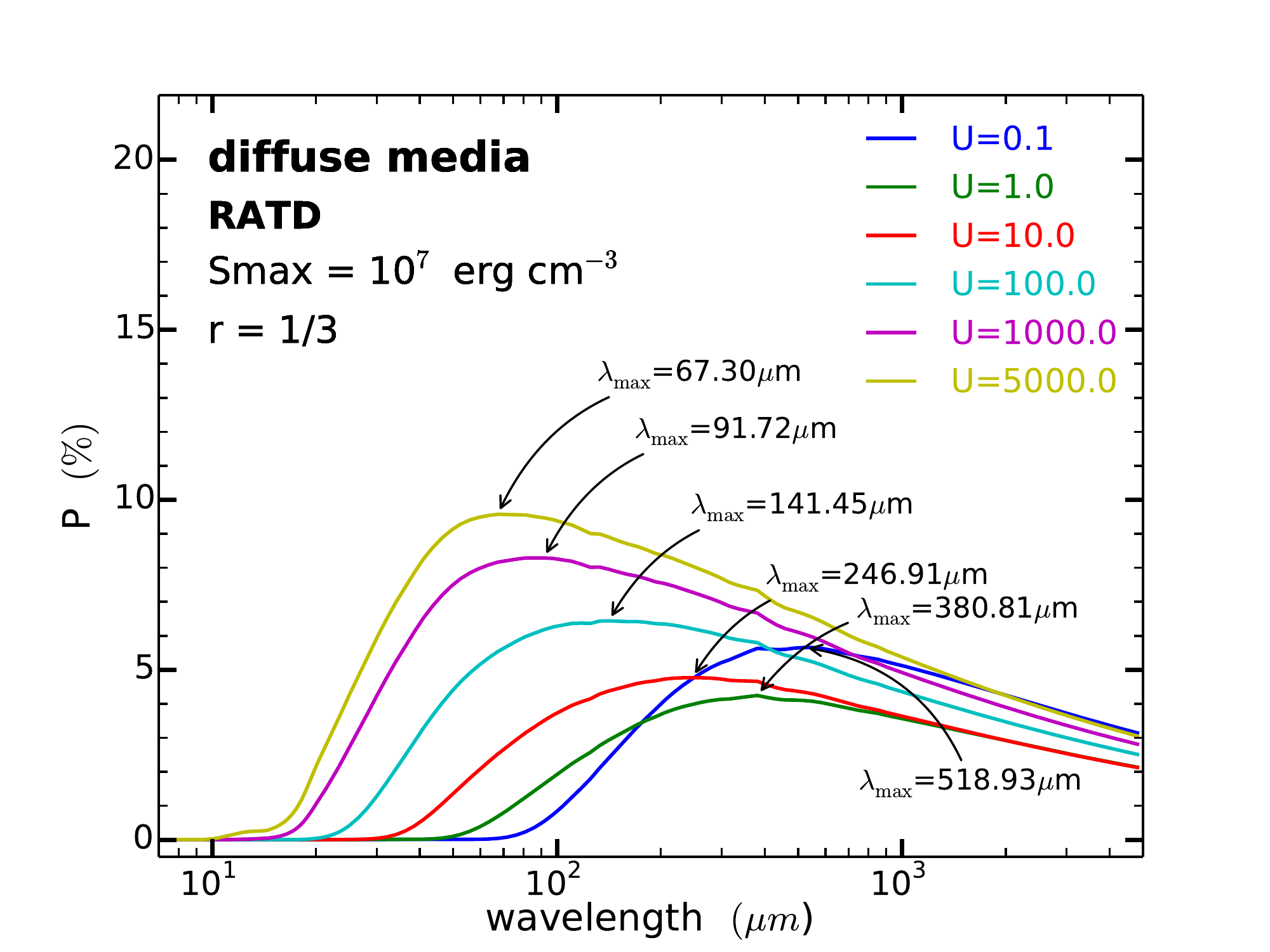}
\caption{Polarization spectrum of thermal emission from aligned grains by RATs with axial ratio $r=1/3$ in the diffuse medium with various radiation field strengths, assuming no grain disruption (left panel) and with disruption by RATD (right panel). The tensile strength $S_{\max}=10^{7}\erg\cm^{-3}$ is considered.} 
\label{fig:PemD}
\end{figure*}

Figure \ref{fig:PemD} shows the polarization spectrum of thermal emission from dust grains aligned by RATs in the absence of RATD (left panel) and presence of RATD (right panel) for prolate grains of axial ratio $r=1/3$, assuming the tensile strength $S_{\max}=10^{7}\erg\cm^{-3}$. 

In the absence of RATD (left panel), the maximum polarization increases with increasing the radiation strength $U$ as a result of enhanced alignment of small grains (see Figure \ref{fig:Afunc}). The peak wavelength ($\lambda_{\max}$) of the polarization spectrum moves toward short wavelengths as $U$ increases, but their spectral profiles remain similar. When the RATD mechanism is taken into account, the polarization degree for $U\gtrsim 1$ is essentially lower than the case without RATD due to the removal of large grains by RATD (see Table \ref{tab:DustSize}). Moreover, the peak polarization degree decreases as the radiation strength increases from $U=0.1$ to $U=1.0$. 

\begin{figure*}
\includegraphics[scale=0.45]{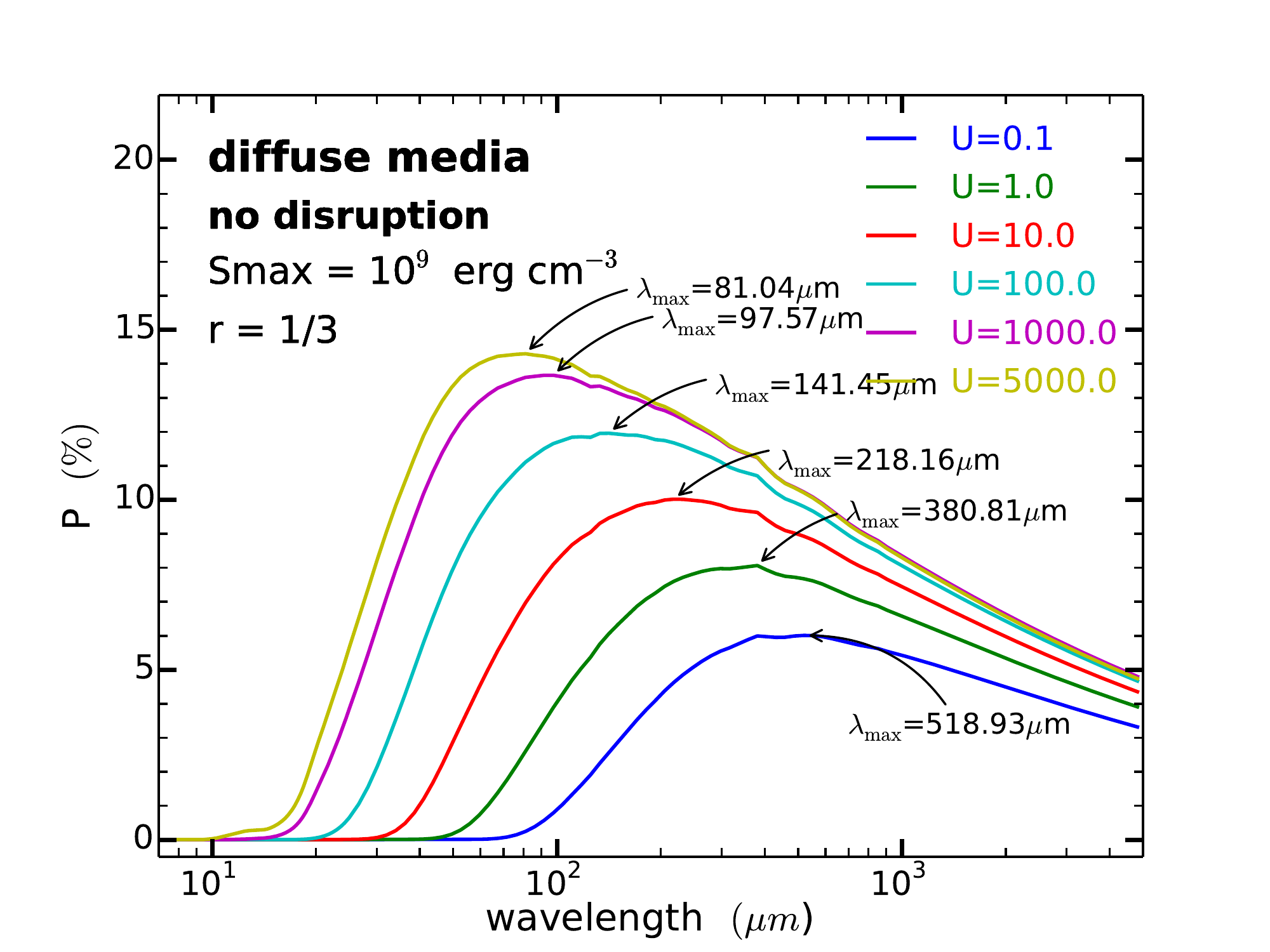}
\includegraphics[scale=0.45]{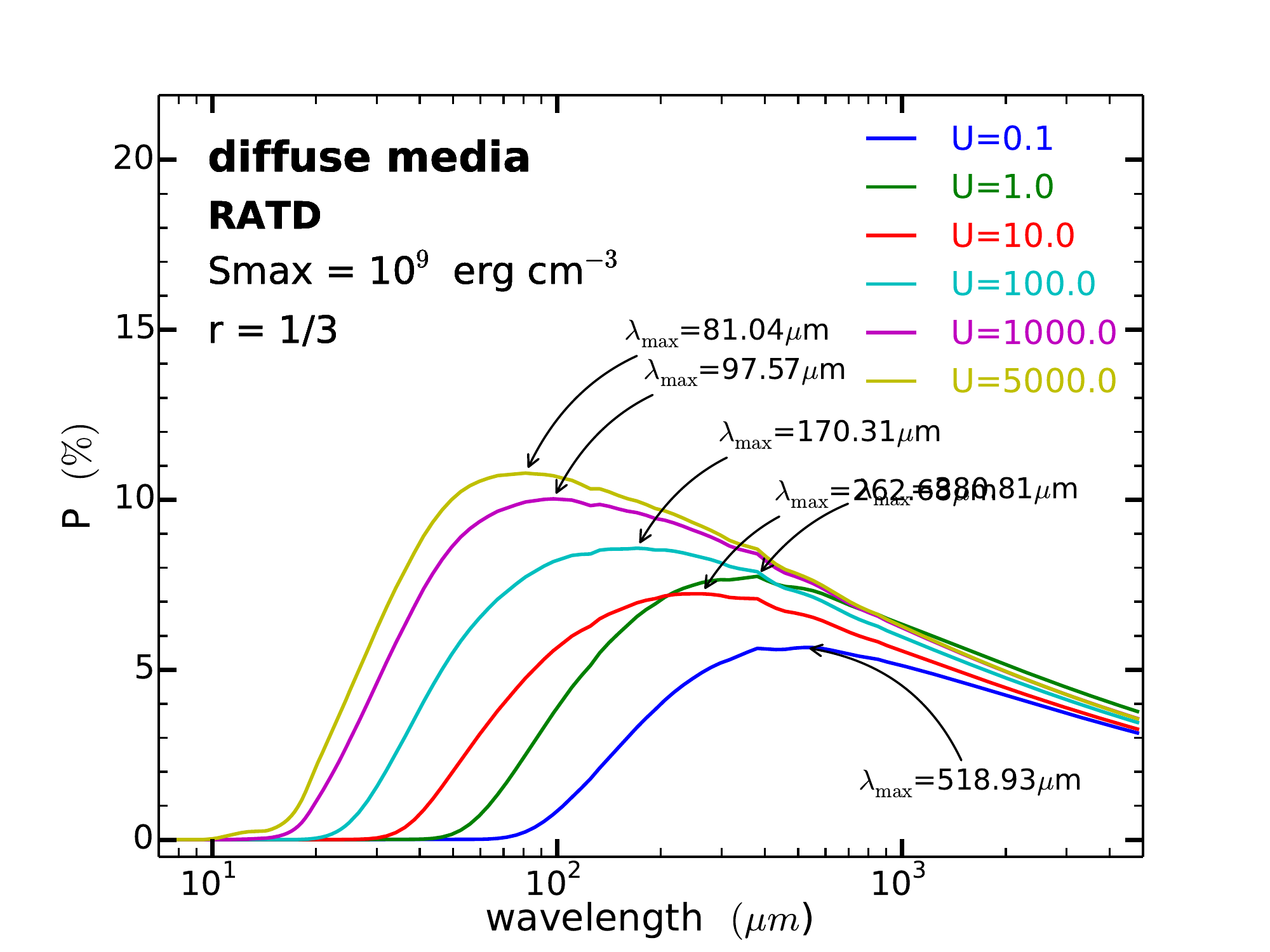}
\caption{Same as Figure \ref{fig:PemD} but for a higher tensile strength of $S_{\max}=10^{9}\erg\cm^{-3}$.}
\label{fig:PemD_S1e9}
\end{figure*}

Figure \ref{fig:PemD_S1e9} shows the results but for dust grains having a higher tensile strength (i.e., $S_{\max}=10^{9}\erg\cm^{-3}$). The similar trend is observed, but the peak polarization increases for $U=0.1-1$ and then decreases as the radiation strength increases from $U=1$ (green line) to $U=10$. The reason is that the disruption requires a higher radiation strength than for grains with lower $S_{\max}$.

\begin{figure*}[h]
\includegraphics[scale=0.45]{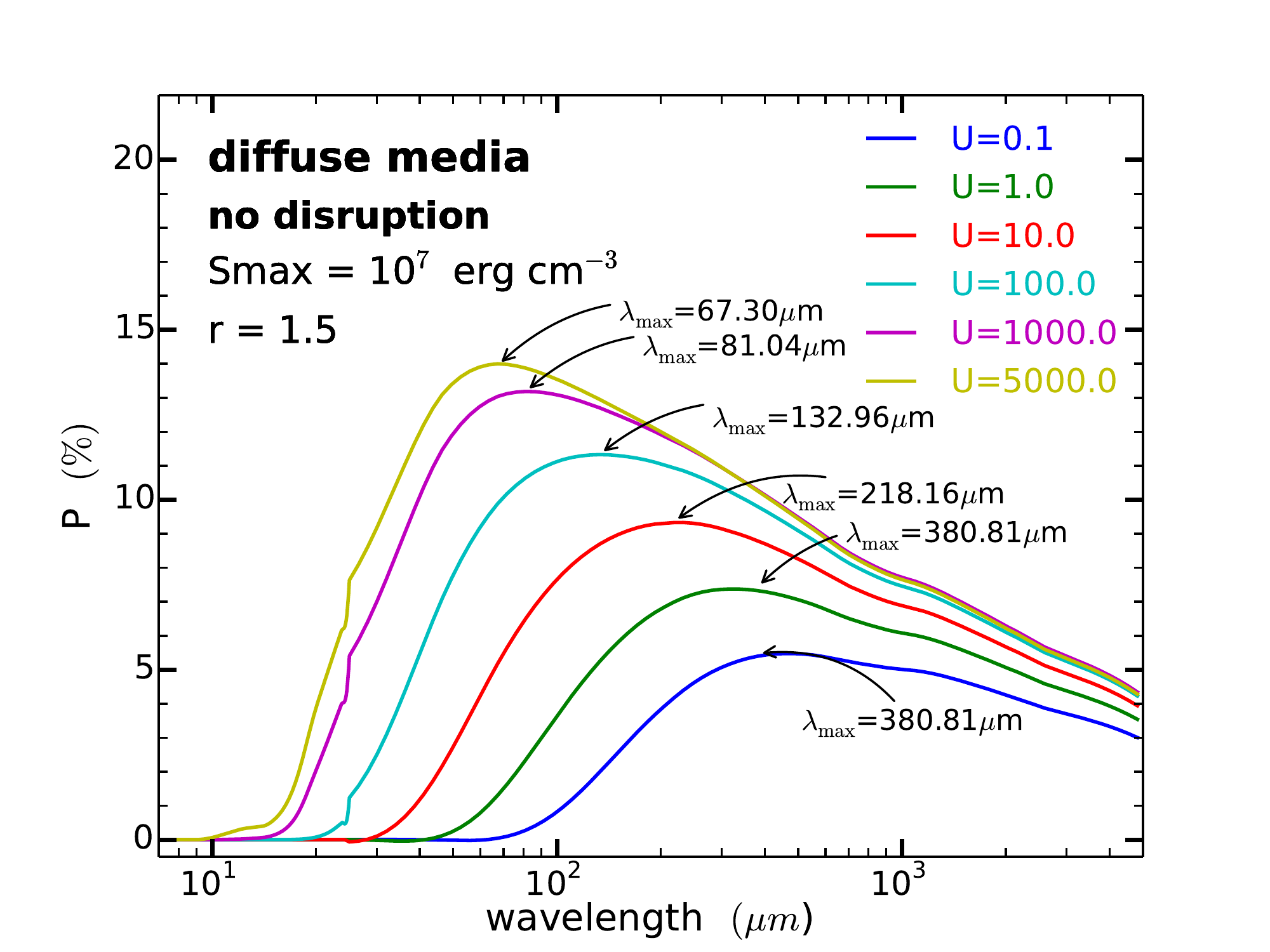}
\includegraphics[scale=0.45]{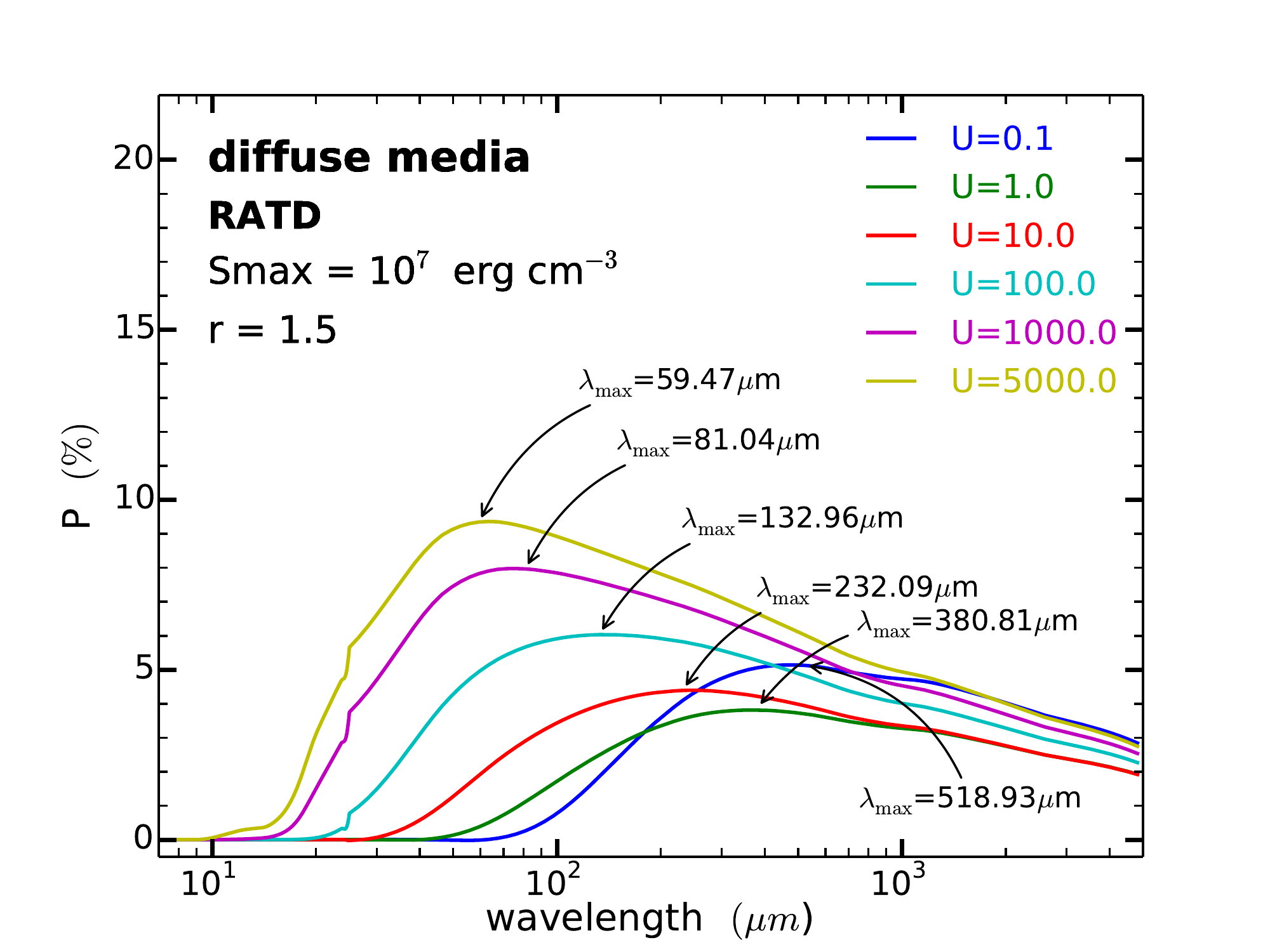}
\caption{Polarization spectrum of thermal emission from aligned grains by RATs with axial ratio, $r=1.5$, in the diffuse ISM with various radiation strengths for two cases without RATD (left panel) and with RATD. The tensile strength $S_{\max}=10^{7}\erg\cm^{-3}$ is considered.}
\label{fig:PemDr15}
\end{figure*}

Figure \ref{fig:PemDr15} shows similar results as Figures \ref{fig:PemD}, but for oblate grains of axial ratio $r=1.5$. It shows that increasing the axial ratio of grains, $r=1.5$, results in shorter peak wavelength due to efficient alignment of elongated dust grains, but the shape of polarization curves is not influenced strongly even in the case for taking account of dust grain disruption. 

\subsection{Polarization spectrum for molecular clouds}
\begin{figure*}
\includegraphics[scale=0.45]{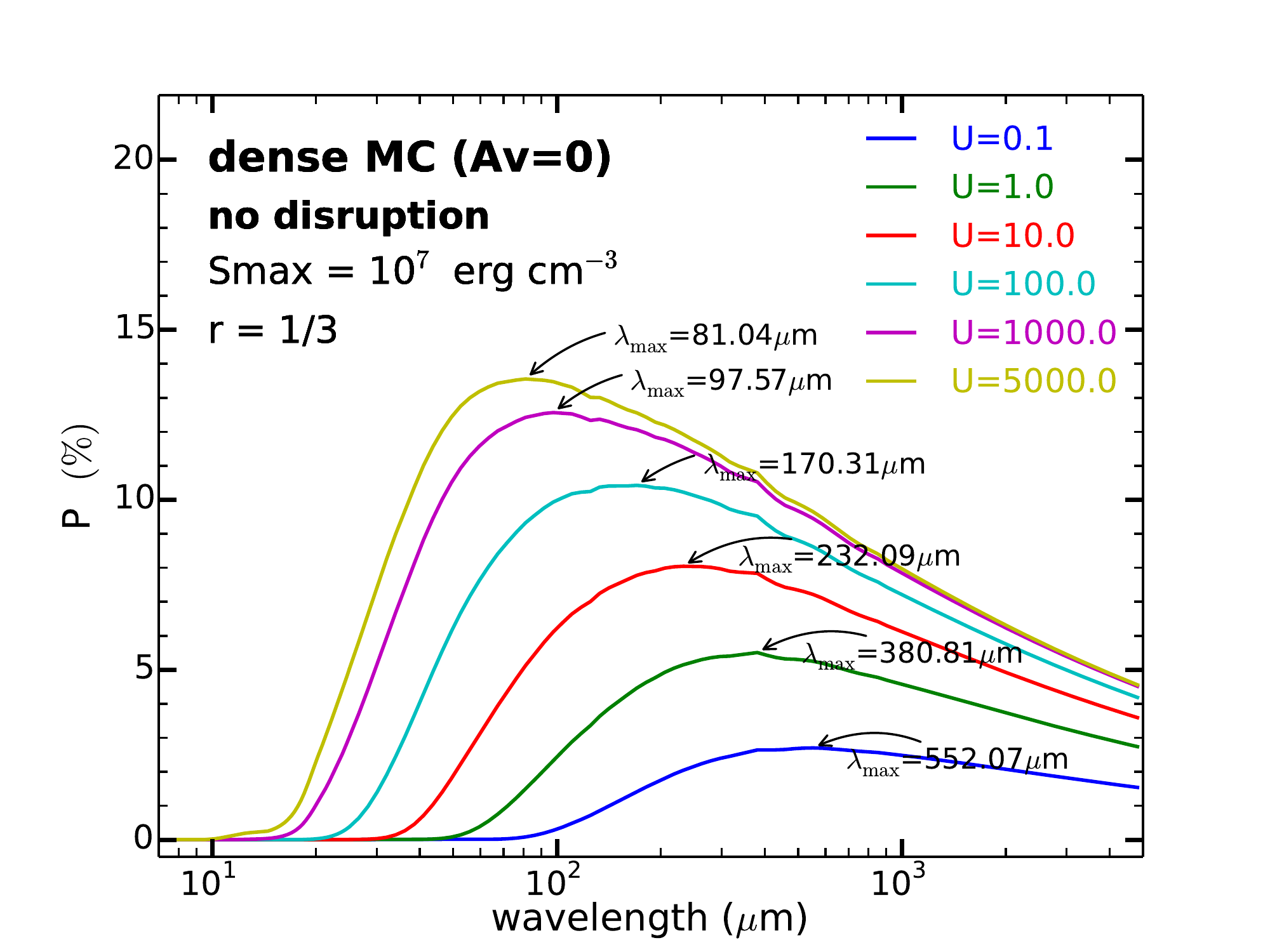}
\includegraphics[scale=0.45]{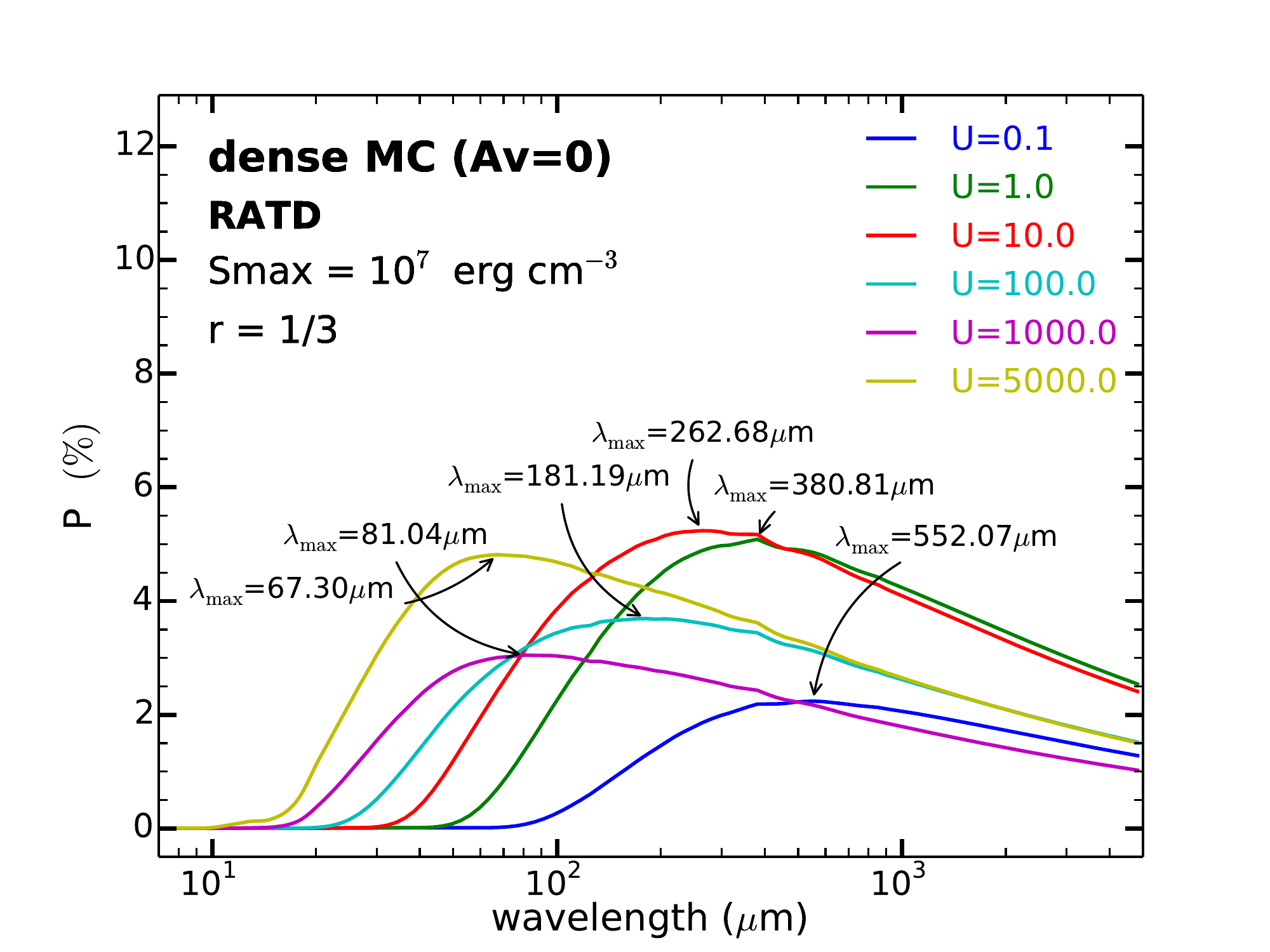}
\caption{Polarization spectrum of thermal dust emission with axial ratio of $1/3$ and tensile strength of $10^9\erg\cm^{-3}$ in a molecular cloud where a star is located in the center of: using grain size distribution including constant maximum size of dust grains aligned by radiation field (left panel), and including aligned grain size less than disruption size (right panel). Polarization spectrum changes with different $U$.}
\label{fig:PemG}
\end{figure*}

Figure \ref{fig:PemG} shows the polarization spectrum obtained for aligned grains in a MC, assuming $S_{\max}=10^{7}\erg\cm^{-3}$. As seen, the polarization degree first increases from $U=0.1$ to $U=10$ and then it falls between $U=10$ and $U=100$ due to the disruption of large grains via the RATD mechanism. Note that for MCs of higher gas density (i.e., faster rotational damping), the rotational disruption occurs at $U\sim 10$ because the required radiation strength must be higher than for the diffuse ISM, assuming the same tensile strength.

\begin{figure}
\includegraphics[scale=0.45]{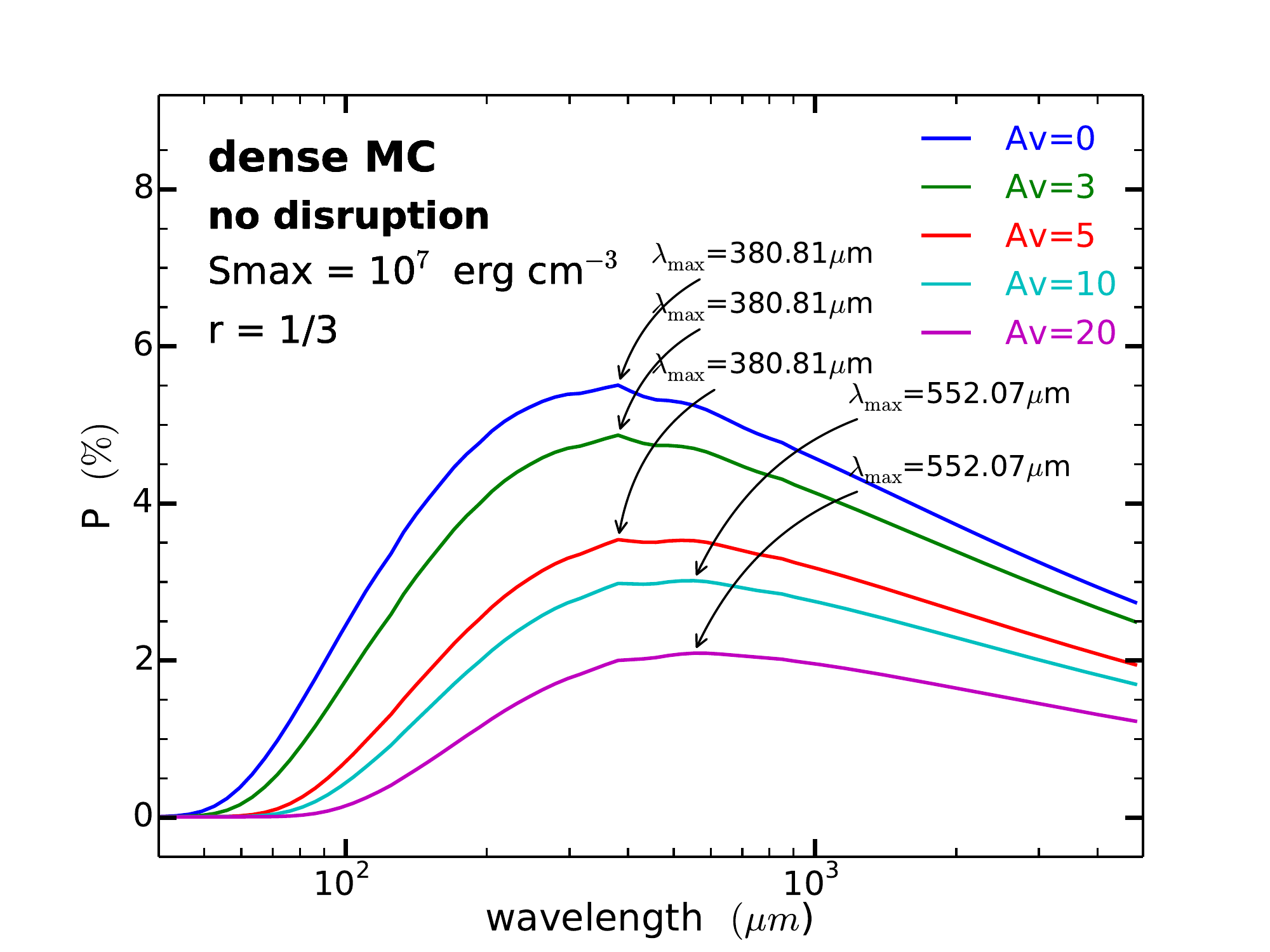}
\caption{Polarization spectrum due to thermal emission of dust grains with axial ratio of $r=1/3$ and tensile strength of $10^7\erg\cm^{-3}$ in a  molecular cloud. Polarization spectrum changes with different $A_{\rm V}$.}
\label{fig:PemG_Av}
\end{figure}

\subsection{Variation of submm polarization with the radiation field}

%\begin{figure*}
%\includegraphics[scale=0.45]{results/fig14-1.pdf}
%\includegraphics[scale=0.45]{results/fig14-3.pdf}
%\caption{Polarization degree at 850 $\mu$m as a function of the radiation strength $U$ for two cases without RATD (solid lines) and with RATD (dashed lines), assuming the diffuse ISM (left panel) and a MC (right panel).  The corresponding grain temperature is shown on the upper horizontal axis.}
%\label{fig:Pem850DG}
%\end{figure*}

To see in more detail how the submm polarization degree changes with $U$ and grain temperature $T_{d}$, we calculate the polarization degree at $\lambda=850\mu$m (P$_{\rm 850}$) using our results in the previous section. Grain temperature is simply estimated from $U$ using the formula $T_{d}=16.4a_{-5}^{1/15}U^{1/6}$ for silicate grains (see \citealt{Draine:2011}).

%Figure \ref{fig:Pem850} shows the change of $P_{850}$ as a function of $U$ ($T_{d}$, top horizontal axis) in the presence of RATD (dashed lines) compared to the results without RATD (solid lines), in the diffuse ISM (left panel), assuming $S_{\max}=10^{7}\erg\cm^{-3}$. In the absence of RATD, the polarization degree increases gradually as the radiation strength increases from $U=0.1$ to $\sim 100$ and changes slowly beyond that. In the presence of RATD, the polarization degree first decreases with increasing $U$ from $U=0.1$ due to the decrease of large grains by rotational disruption. After that, it starts to rise because the abundance of aligned grains increases due to smaller $a_{\rm align}$.

%The right panel shows the similar results, but by aligned grains in a MC. The similar trend is observed. One prominent difference is that $P_{850}$ first increases with the radiation from $U=0.1$ to $U\sim 1$ and then declines rapidly due to RATD. Such a difference arises from the fact that due to higher gas number density in MCs, the rotation rate spun-up by RATs is smaller (see Eq. \ref{eq:wRAT}), resulting in a higher $U$ required for RATD. Similarly, in the middle panel, we see a rapid decrease of the polarization degree with the radiation strength from $U=1.0$ to $U=10$ in a medium with n$_{\rm H}$=100cm$^{-3}$. Compared with the case of MCs, a rapid decrease begins at a lower radiation strength, which arises from the fact that lower gas density facilitates faster rotation of grains by RATs due to weaker rotational damping.

\begin{figure*}
\includegraphics[scale=0.45]{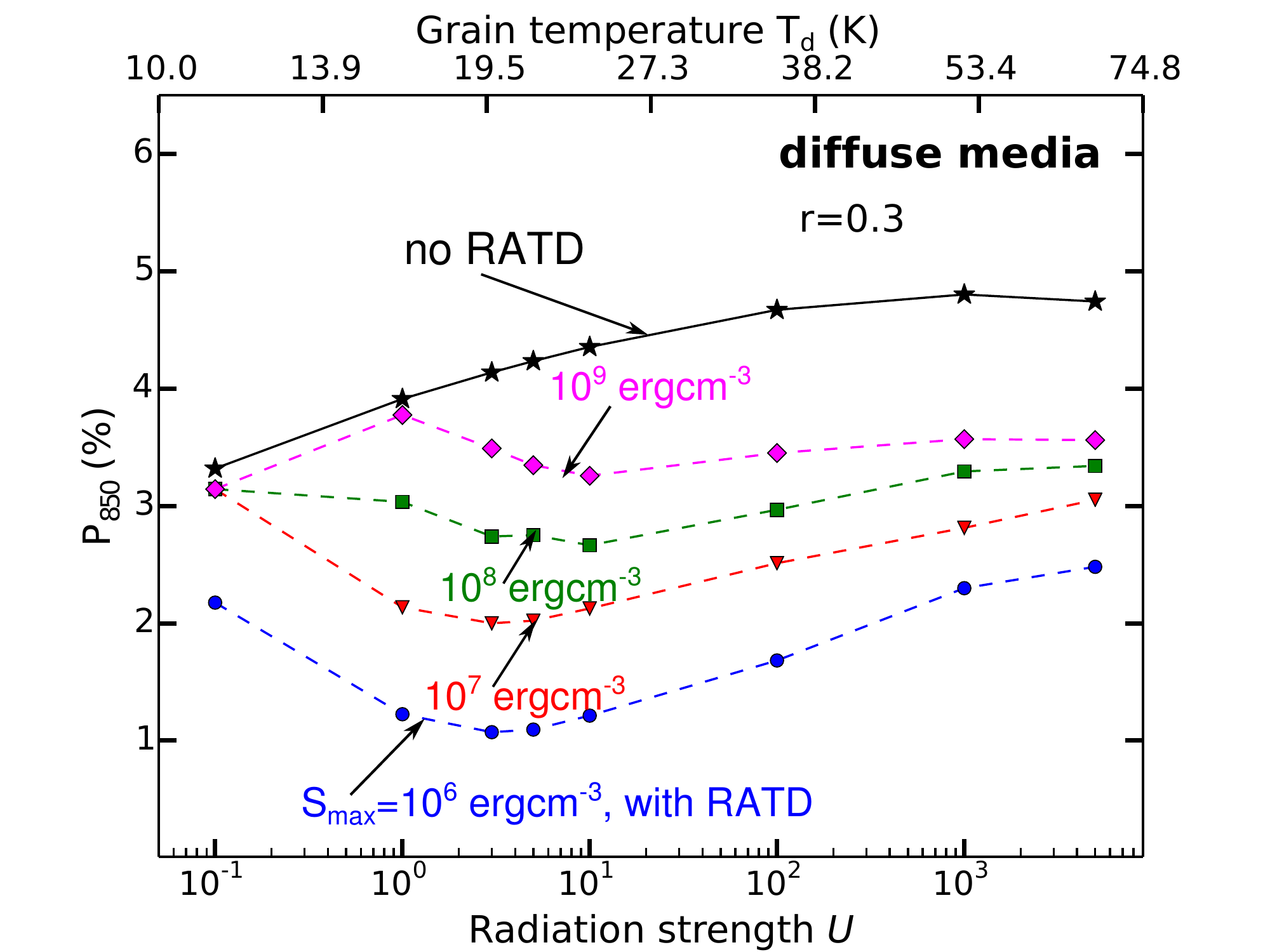} 
\includegraphics[scale=0.45]{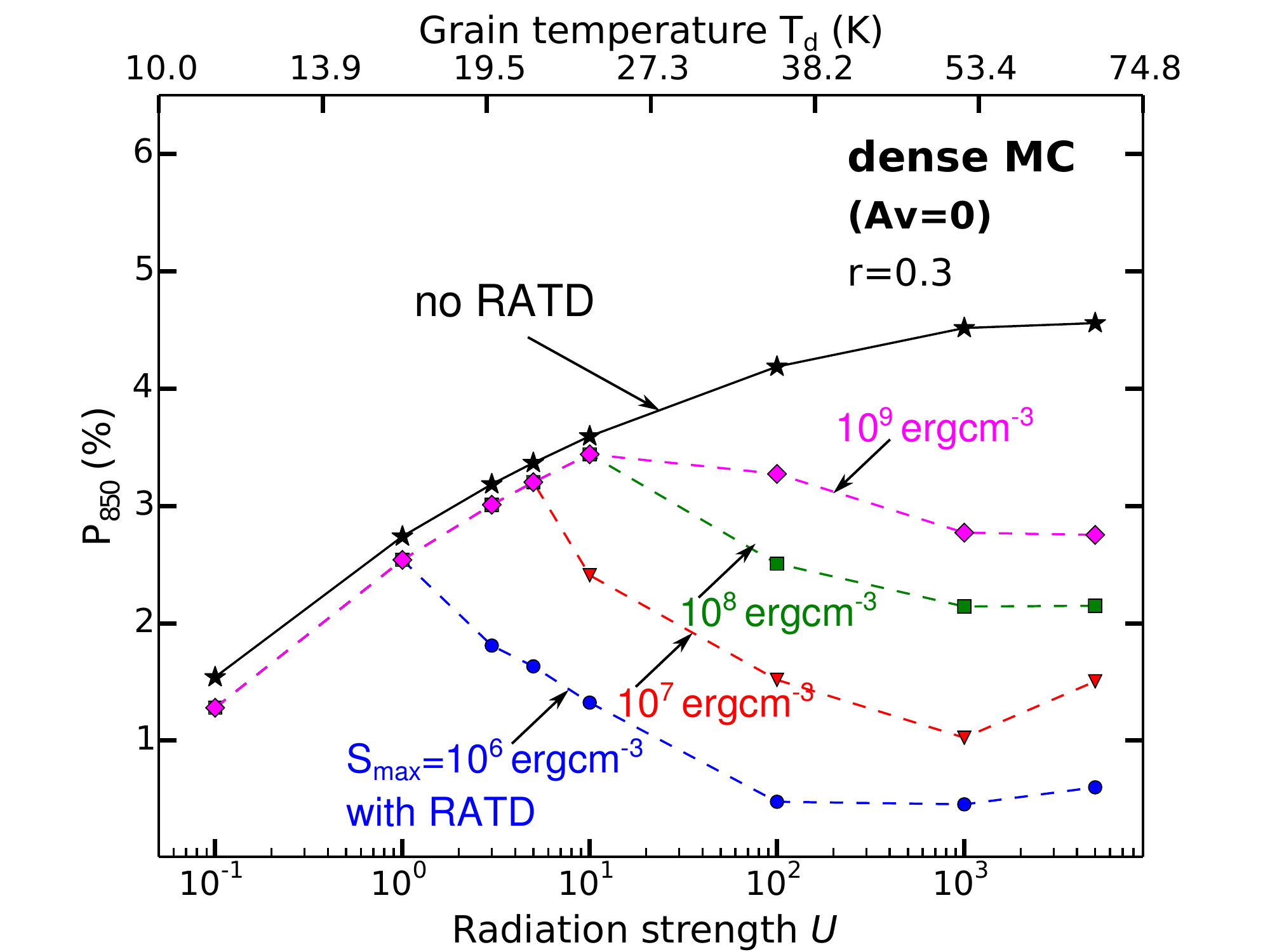} 
\caption{Polarization degree at 850$\mu$m for the different radiation strength ($U$) or grain temperature ($T_d$, top horizontal axis) for two cases, without RATD (solid lines) and with RATD (dashed lines), assuming the different tensile strength of grains in the diffuse ISM (left panel) and a MC (right panel). Grains with axial ratio of $r=1/3$ are considered.}
\label{fig:Pem850}
\end{figure*}

In Figure \ref{fig:Pem850}, we show the variation $P_{850}$ with the radiation strength $U$ or grain temperature $T_d$, calculated for grains in the diffuse ISM (left panel) and molecular clouds (right panel), assuming a wide range of the tensile strength $S_{\max}=10^{6}-10^{9}\erg\cm^{-3}$. The black line shows the results when the RATD is not taken into account. In contrast to the increase of $P_{850}$ with $U$ in the absence of RATD, the polarization $P_{850}$ in the diffuse ISM does not change considerably when the radiation strength increases between $3-100$ when RATD is accounted for. This is because of the compensation of the shift of polarization toward short wavelengths due to lower $a_{\rm align}$ and the increase of the polarization. Indeed, in the case of high tensile strength ($S_{\max}\ge 10^{8}\erg\cm^{-3}$), we cannot expect the overall increase of the polarization degree with $U$, but rather, the variation in the wavelength dependence polarization. When $U\le 1$ ($T_{d}<16.4\K)$ for $S_{\max}=10^{9}\erg\cm^{-3}$, the polarization $P_{850}$ increases as U increases. The peak of polarization $P_{850}$ moves to a smaller radiation strength or lower grain temperature for a smaller $S_{\max}$. 

Figure \ref{fig:Pem850} (right panel) show the similar results but for a MC. The amplitude of the polarization variation with $U$ due to RATD is larger for the MC (see right panel of Figure \ref{fig:Pem850}). Within the RAT paradigm, the wide amplitude of the change for the MC is understood because for a high gas number density $n_{\H}$, the RATD requires a higher radiation strength to be effective. So, in the case of high tensile strength ($S_{\max}\ge 10^{8}\erg\cm^{-3}$) for dense MC, when $U$ increases from $U=0.1$, the polarization increases until $U\sim10$ and then decreases due to RATD. The larger pumping range raises the peak of the polarization as seen in the right panel of Figure \ref{fig:Pem850}.

\begin{figure*}
\includegraphics[scale=0.45]{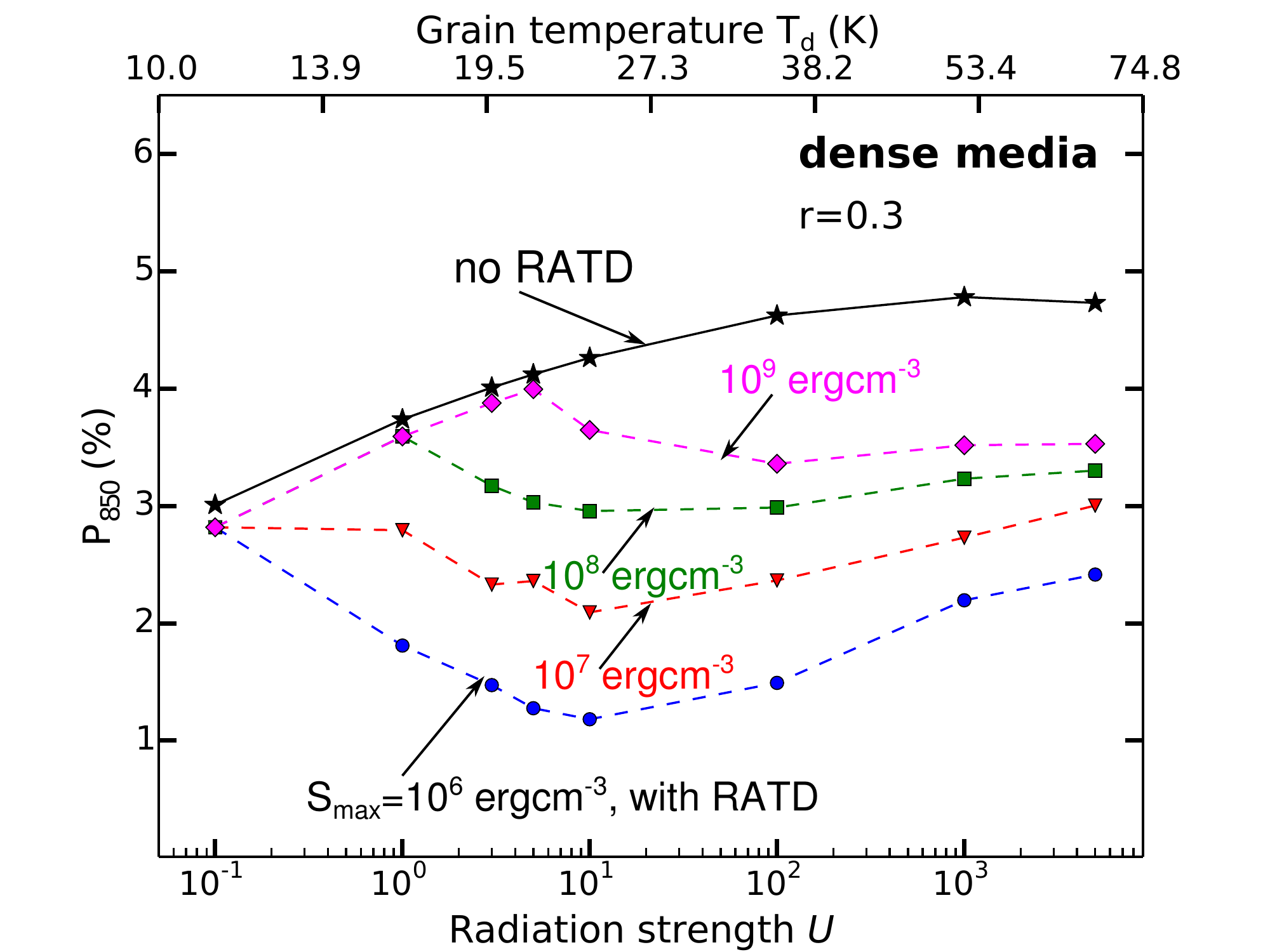} 
\includegraphics[scale=0.45]{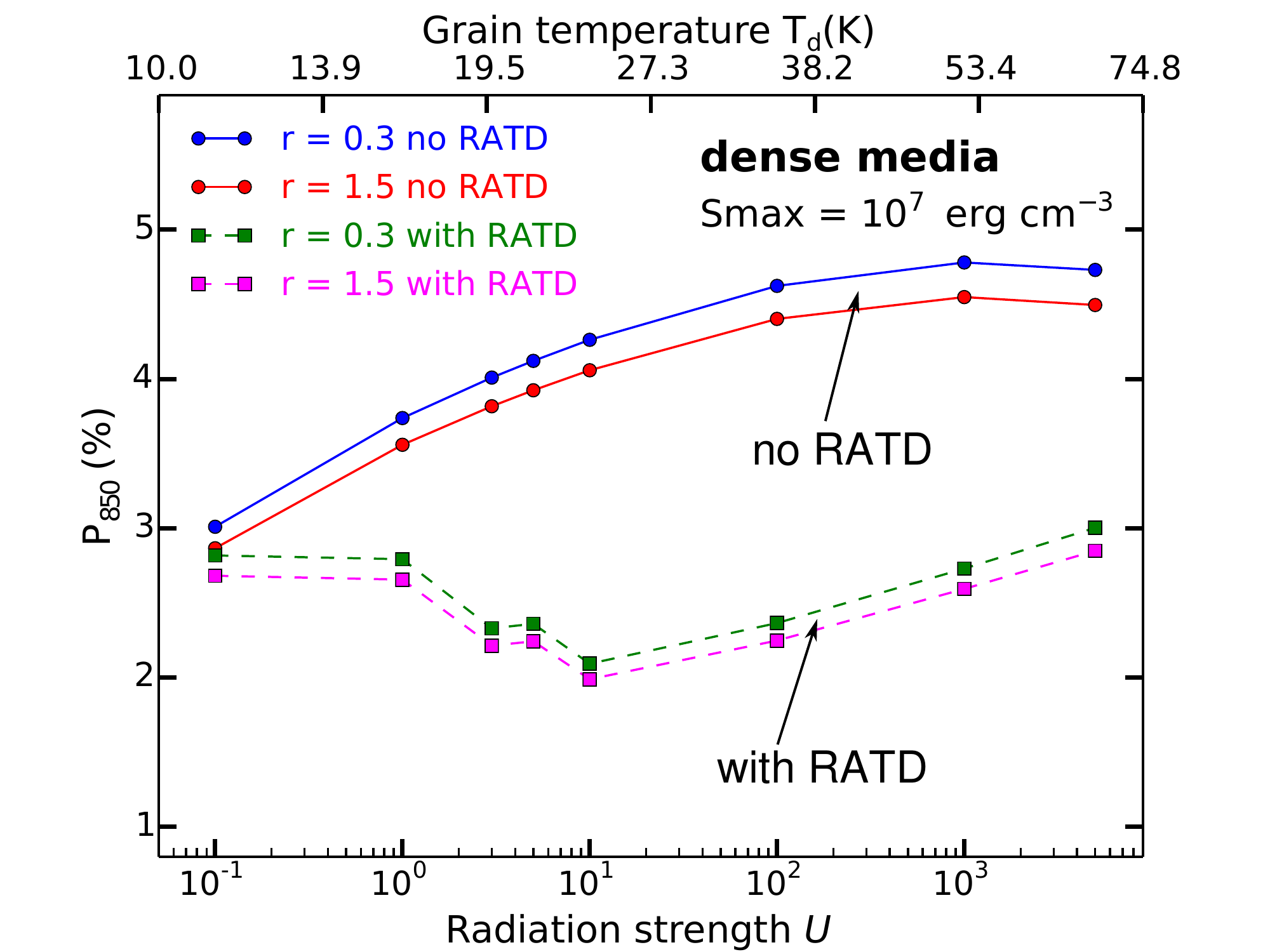}
\caption{Left panel is the same as Figures \ref{fig:Pem850} but for a translucent cloud with density $n_{\H}=100\cm^{-3}$. Right panel shows the variation of $P_{850}$ for two grain shapes of axial ratio $r=0.3$ and $r=1.5$.}
\label{fig:Pem850_n100}
\end{figure*}

Figure \ref{fig:Pem850_n100} (left panel) shows the results for a translucent cloud with gas density between the diffuse ISM and dense MC. The similar trend is observed, but the critical strength where $P_{850}$ starts to decrease is larger than the diffuse ISM, but smaller than the MC. We also study the variation of $P_{850}$ with $U$ for grains of axial ratio $r=1.5$ in the right panel and find the similar trend.

% ========== Discussion ========== 

\section{Discussion}\label{sec:discussion}
\subsection{Physical forward modeling of multi-wavelength polarization}\label{sec:multi-lamb}
The polarization spectrum closely depends on the grain size distribution and alignment degree of dust grains. Both the grain size and alignment are expected to change with the local environments. Inverse modeling of observational data (e.g., \citealt{2009ApJ...696....1D}; \citealt{2018A&A...610A..16G}) is a useful technique to derive the average property of dust grains. 

In this paper, we focus on the variation of the local radiation strength $U$ and perform forward modeling of multi-wavelength dust polarization from UV-optical-NIR (starlight polarization) to far-IR (polarized emission) to predict how the polarization spectrum changes with increasing $U$ from the standard ISRF. We simultaneously treat grain alignment and disruption by RATs. The grain size distribution is modeled consistently using the RATD mechanism, which changes with the strength of the radiation field, as shown in Section \ref{sec:Align}. Our modeling results show that when the radiation strength $U$ increases, the polarization spectrum in general shifts to short wavelengths (see Figures \ref{fig:PabsDiff}-\ref{fig:PabsDiffr15} for starlight polarization and Figures \ref{fig:PemD}-\ref{fig:PemDr15} for polarized thermal emission). At the same time, the maximum polarization degree of starlight as well as thermal dust emission also increases with increasing $U$.

Thanks to the RATD effect, for the first time, we can study the dependence of the interstellar polarization spectrum on the mechanical properties of dust, characterized by the tensile strength $S_{\max}$. For a given radiation field, our results show that the polarization spectrum depends crucially on $S_{\max}$ because the RATD determines the upper cutoff of the grain size distribution.
% v-shape

Previously, \cite{2018A&A...610A..16G} modeled the dust polarization spectrum for different local radiation strength of $U=0.1$ to $U=10^{3}$ using the best-fit alignment function (model D) obtained from fitting the average full-sky Planck data. This model does not take into account the variation of grain alignment efficiency with $U$. As $U$ increases, the polarization spectrum shifts to short wavelengths, but the peak polarization slightly changes.

\subsection{Towards constraining grain internal structures with observational data}\label{sec:planck}

In Figure \ref{fig:Pem850}, we have shown that in the absence of grain disruption by RATD, the polarization at $850\mum$, denoted by $P_{850}$, increases monotonically with the radiation intensity (i.e., grain temperature) over the considered range of $U$. The absence of RATD is equivalent to the situation where grains are made of ideal material without impurity such that the tensile strength is as high as $S_{\max}\sim 10^{11}\erg\cm^{-3}$ (e.g., diamonds). However, when the RATD effect is taken into account for grains made of weaker material ($S_{\max}\lesssim 10^{9}\erg\cm^{-3})$, the variation of the polarization degree $P_{850}$ with $U$ depends closely on the tensile strength. The general trend is that $P_{850}$ first increases from a low value of $U$ and then decreases when $U$ becomes sufficiently large. The critical value $U$ at the turning point is determined by the value $S_{\rm max}$ and local gas density $n_{\H}$ that controls the grain disruption size $a_{\rm disr}$ according to RATD.

\cite{2018arXiv180706212P} performed a detailed analysis of the variation of $P_{850}$ with the radiation field using {\it Planck data}. The authors discovered that $P_{850}$ first increases with increasing grain temperature from $T_{d}\sim 16-19\K$ and then drops as the dust temperature increases to $T_{d}\gtrsim 19\K$. Such an unusual $P_{850}-T_{d}$ relationship cannot be reproduced if large grains are not disrupted (i.e., RATD is not taken into account), as shown in Figures \ref{fig:Pem850} and \ref{fig:Pem850_n100}. Moreover, the observed trend is, in general, consistent with our model with RATD, but grains have the tensile strength of $S_{\max}\lesssim 10^{9}\erg\cm^{-3}$. This range of tensile strength favors a composite internal structure of grains over the compact one.

We also note that the polarization degree of polarized thermal emission obtained from our model is lower than predicted by \cite{2018A&A...610A..16G}. The difference perhaps arises from the fact that we adopt a power-law size distribution instead of using the best-fit size distribution to the Planck data by \cite{2018A&A...610A..16G}. However, we focus on the overall polarization spectrum with the varying radiation strength instead of fitting to the observational data.

\subsection{Comparison to the optical polarization of SNe Ia}\label{sec:SN}
Due to extinction by aligned grains, the starlight is polarized, and the degree of polarization varies with the wavelength. In general, the maximum polarization occurs at the peak wavelength of $\lambda_{\max} \sim 0.55 \mu$m and declines on both sides of the peak. Following \cite{1973IAUS...52..145S}, the polarization curve of starlight can be described by an empirical formula (namely Serkowski law):

\begin{equation}
P(\lambda) = P_{\max}\exp\left(-K\ln^2(\lambda/\lambda_{\max})\right)
\label{eq:serkowski}
\end{equation}
where $P_{\max}$ is the maximum of polarization, $\lambda_{\max}$ is the peak wavelength at $P_{\max}$, and $K$ is a parameter that characterizes the "width" of the polarization profile (see more in Section\ref{sec:SN}). 

\begin{figure}
\includegraphics[scale=0.5]{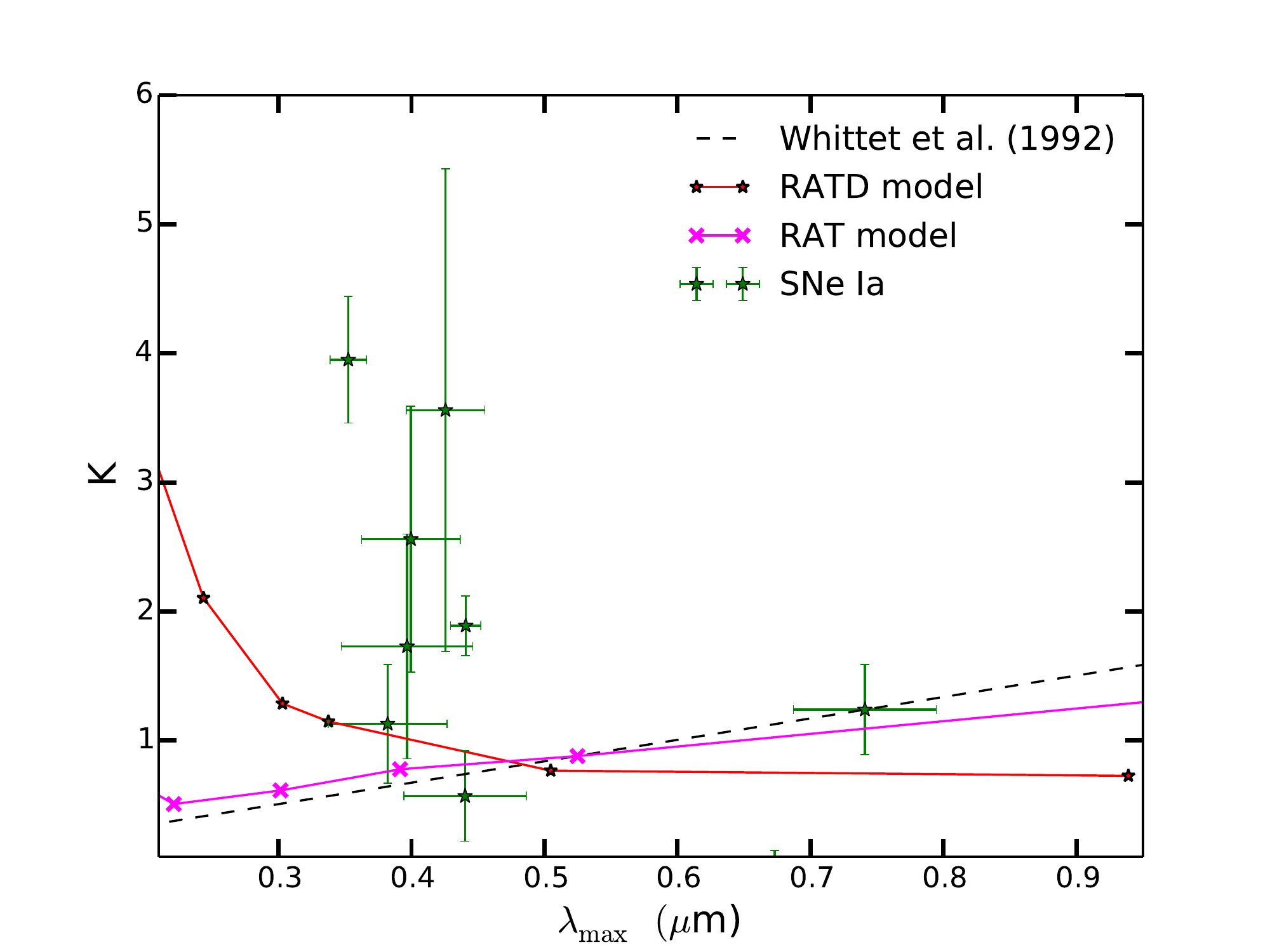} 
\caption{Extinction sight-lines in the $\lambda_{\max}$-K plane. A magenta solid line and a red solid line show the calculation adopted RAT and RATD, respectively, for dust grains with $S_{\max}$ of 10$^7$ erg/cm$^3$ and axial ratio of $r=1/3$ in the diffuse ISM. The dashed black line traces the relation in \cite{1992ApJ...386..562W}. Points are the observational data: green stars are the samples of SNe Ia from \cite{2015A&A...577A..53P} and \cite{2017ApJ...836...88Z}.}
\label{fig:lamb-K}
\end{figure}
% The red line represents samples of SNe Ia from \citealt{2015A&A...577A..53P} and \citealt{2017ApJ...836...88Z}.

Polarimetric observations of SNe Ia are an excellent test for our theoretical prediction of dust polarization. The polarization curve has correlation between Serkowski parameters K and $\lambda_{\max}$. The "width" parameter K, in Equation (\ref{eq:serkowski}), is linearly correlated with $\lambda_{\max}$ as $K=c_1\lambda_{\max}+c_2$, where $c_1$ and $c_2$ are the average values of the slope and intercept in K-$\lambda_{\rm max}$ plane. For standard K-$\lambda_{\max}$ relationship, the current best value for $c_1$ and $c_2$ are $1.66\pm 0.09$ and $0.01\pm 0.05$ respectively (see \citealt{2003dge..conf.....W}). The smaller $K$ shows broader profile in polarization curve. The correlation of $\lambda_{\max}$ and $K$ is shown in Figure \ref{fig:lamb-K} where the standard relationship by \cite{1992ApJ...386..562W} is presented in the dashed black line. The left panel of Figure \ref{fig:PabsDiff} shows broader curve as radiation strength becomes higher. From the curve, we calculate $\lambda_{\max}$ and derive the $K$ value by fitting the Serkowski law to the calculated polarization curve. Its correlation is shown as a magenta line in the Figure \ref{fig:lamb-K}. Our calculation for the model of RAT alignment is consistent with the standard relationship. In the right panel of Figure \ref{fig:PabsDiff}, on the other hand, we find that the curve becomes narrower and $\lambda_{\max}$ gets shorter for a higher radiation strength. This result of RATD model is consistent with the study of \cite{2018A&A...615A..42C}. 

Peculiar polarization data observed toward SNe Ia by \cite{2018A&A...615A..42C} show that the $K$ parameter does not follow the standard relationship. In order to see if the RATD mechanism can explain SNe Ia polarization data, we calculate the $K$ parameter and $\lambda_{\max}$ using the polarization curves from Section \ref{sec:Pabs}. The red line is our samples calculated in consideration of RATD mechanism with aligned grains which has axial ratio, $r=1/3$, at $S_{\max}=10^7\erg\cm^{-3}$. A least-square fit to our samples has slope and intercept $c_1= -8.5$ and $c_2= 6.5$. The relationship in our calculation is significantly different from the relation for dust grains in the Taurus region (\citealt{1992ApJ...386..562W}), while it is similar with samples of SNe \rom{1}a. The strong radiation from a hot source like as SNe \rom{1}a can disrupt large grains and form small grains. These small grains are aligned by radiation and produce large K at shorter $\lambda_{\max}$, representing negative correlation between K and $\lambda_{\max}$. 

% SN \rom{1}a with anomalous extinction sightlines observed from \citealt{2015A&A...577A..53P} and \citealt{2017ApJ...836...88Z} (red line)

%\subsection{Observational testing of grain alignment and disruption by RATs in star-forming regions}

Finally, our calculations assumed the grain size distribution with the standard slope of $\alpha=-3.5$. In principle, the disruption of large grains by RATD enhances the abundance of small grains, so that the size distribution may be steeper than the standard value as long as RATD occurs (\citealt{Giangetal:2019}). To see how the slope affects the polarization polarization, we repeat our calculations for $\alpha=-4$. We find that the obtained results are slightly different from results shown in Figure \ref{fig:PabsGMC_Av}.

% ========== Summary ========== 

\section{Summary}\label{sec:summary}

In this paper, we have performed physical modeling of multi-wavelength polarization by aligned grains for the different radiation fields. Our main results are summarized as follows:

\begin{enumerate}

\item Using the RAT alignment and RATD theory, we obtain the grain alignment function and size distribution of dust grains for the ISM with various radiation fields and model the polarization of starlight and polarized thermal emission by aligned grains.

\item For the diffuse medium, we find that the polarization spectrum of starlight is shifted to the shorter wavelength due to the enhancement of small grains when the radiation intensity increases. At the same time, the optical/NIR polarization is reduced due to the disruption of large grains into smaller ones.

\item For polarized thermal emission, we find that the peak polarization increases but the peak wavelength decreases with increasing radiation strength $U$ due to enhanced alignment of small grains. This prediction can be tested with observations such as by SOFIA/HAWC+.

\item In the absence of RATD, we find that the submm polarization degree at 850 $\mu$m ($P_{850}$) increases with increasing grain temperature ($T_{d}$) until $T_{d}\sim 50\K$. However, when taking into account RATD, we find that the variation of the polarization degree with the radiation strength depends on the tensile strength of grain materials. 

\item Comparing our predictions of $P_{850}-T_{d}$ with the results from \cite{2018arXiv180706212P} using {\it Planck} data, we find that grain disruption must occur in order to reproduce the observed non-monotonic increase of $P_{850}$ with $T_{d}$. This suggests that interstellar grains unlikely to have a compact structure with very high tensile strength but perhaps a composite structure.

\item Based on our results, we suggest that an important way to test RAT theory and RATD is to observe polarization toward star-forming regions. This is a complementary to the traditional way to test RAT for starless cores.

\item Our models of starlight polarization for high radiation intensity with RATD find that the $K-\lambda_{\max}$ does not follow a standard relationship observed for the average ISRF. However, this predicted trend qualitatively agree with observations toward SNe Ia.

\end{enumerate}

\acknowledgments
We are grateful to A. Lazarian for his warm encouragements. We thank V. Guillet for sharing with us the data of cross-section of dust grains used in their paper. This work was supported by the National Research Foundation of Korea (NRF) grants funded by the Korea government (MSIT) through the Basic Science Research Program (2017R1D1A1B03035359) and Mid-career Research Program (2019R1A2C1087045).

%\appedgemtndix

%--------------adding references-----------------------------------
\bibliographystyle{/Users/thiemhoang/Dropbox/Papers2/apj}
% or other styles: mcbride,plain, abbrv, acm, alpha, apalike, apj
\bibliography{/Users/thiemhoang/Dropbox/Papers2/cites_paperApJ,/Users/thiemhoang/Dropbox/Papers2/cites_Books}
%\bibliography{ms.bbl}

\clearpage

%===========
%APPENDIX
%===========
\appendix
\section{Rotational damping process} \label{sec:damping}

%The rotating small grains are damping due to gas-grain or plasma-grain interaction, infrared emission and radio emission. 

The characteristic rotating damping time is 

\bea
\tau_{\gas}=\frac{\pi\alpha_{1}\rho a}{3\delta n_{\H}(2\pi m_{\H}kT_{\gas})^{1/2}}\simeq 8.74\times10^{4}\times\frac{\alpha_{1}}{\delta}a_{-5}\hat{\rho}\left(\frac{30\rm{cm}^{-3}}{n_{\rm H}}\right)\left(\frac{100\rm{K}}{T_{\gas}} \right)^{1/2} yr
\ena
where $\delta$ is the drag coefficient of order unity, and $\alpha_1$ is geometric factor. In general, we expect $\delta\sim\alpha_1$.

A rotating dust grain additionally associated with absorption and emission of photons also has the damping time due to thermal emission,

\begin{equation}
\begin{split}
\tau_{\rm em}&=\frac{8\alpha_{1}(\beta+3)}{5}\frac{\zeta(\beta+4)}{\zeta(\beta+3)}\frac{\rho a^{3}}{\hbar^{2}cu_{\rm{rad}}\left\langle Q_{\rm{abs}}\right\rangle}=1.60\times10^{5}\times\frac{\alpha_1 a_{-5}^3\rho_3}{\left\langle Q_{\rm abs}\right\rangle}\left(\frac{u_{\rm ISRF}}{u_{\rm rad}}\right)\left(\frac{T_d}{18\rm K}\right)^2 yr
\end{split}
\end{equation}
where $\zeta(x)$ denotes the Riemann $\zeta$-function and $Q_{abs}$ is absorption efficiency factor defined as

\begin{equation}
\left\langle Q_{\rm abs}\right \rangle\equiv\frac{1}{u_{\rm rad}}\int u_{\lambda}Q_{\rm abs}(\lambda)d\lambda .
\end{equation}

The total rotational damping rate, $\tau_{\rm drag}$, by gas collisions and thermal emission can be written as

\begin{equation}
\tau_{\rm drag}^{-1} = \tau_{\rm gas}^{-1}+\tau_{\rm em}^{-1}=\tau_{\gas}^{-1}\left(1+F_{\rm IR}\right),
\end{equation}
where $F_{\rm IR}=\tau_{\rm gas}/\tau_{em}$.

Once we know RAT efficiency, $Q_{\Gamma}$ (using Equations \ref{eq:RATs} \& \ref{eq:Qgam}), we can obtain the angular velocity of grain rotation as follows:
\begin{equation} 
\omega_{\rm RAT} = \frac{\left|\Gamma_{\rm RAT}\right|}{(2/3)\delta n_{\rm H}(1.2)(8\pi m_{\rm H}kT_{\gas})^{1/2}a^4} \left(\frac{1}{1+\tau_{\rm gas}/\tau_{\rm em}} \right).
\label{eq:omega_rad}
\end{equation} 

%\begin{equation} \label{eq:omega-ratio}
%\left(\frac{\omega_{\rm{rad}}}{\omega_{\rm T}}\right)^{2}=\left[\frac{\Gamma_{\rm rad}}{(2/3)\delta n_{\rm H}(1.2)(8\pi m_{\rm H}kT)^{1/2}a_{\rm eff}^{4}}\right]^{2}\times\left(\frac{8\pi\alpha_{1}\rho a_{\rm eff}^{5}}{15kT}\right)^{2}\left(\frac{1}{1+\tau_{\rm gas}/\tau_{\rm em}}\right)^{2}.
%\end{equation}

% ------------ cross section
\section{Cross Section} \label{sec:cross-section}

When starlight traveling in space encounter dust grains on the line of sight, it induces the extinction and polarization of starlight due to absorption and scattering by dust grains. Assume an oblate spheroidal grin with a radius $a$. The extinction is the sum of absorption and scattering, so that we have $Q_{\rm ext}\equiv Q_{\rm abs} + Q_{\rm sca}$. Here, we define efficiency factors $Q_{\rm abs}$, $Q_{\rm sca}$ by 
\begin{equation} 
Q_j \equiv \frac{C_j}{\pi a^2}
\label{eq:efficiency}
\end{equation}
where j=absorption or scattering.

Since we shall treat light as electromagnetic waves, the cross-section for extinction of waves on dust articles is represented as following:

%Let us consider an oblate spheroidal grain with the symmetry axis $\ba_{1}$ having an effective size $a$, which is the size of an equivalent sphere of the same volume as the grain. A perfectly polarized electromagnetic wave with the electric field vector $\bE$ is assumed to propagate along the line of sight. Let $C_{\ext}(\bE\perp \ba_{1})$ and $C_{\ext}(\bE\| \ba_{1})$ be the extinction of radiation by the grain for the cases in which $\bE$ is parallel and perpendicular to $\ba_{1}$, respectively. For the sake of simplification, we denote these extinction cross sections by $C_{\perp}$ and $C_{\|}$. 

%For the general case in which $\bE$ makes an angle $\theta$ with $\ba_{1}$, the extinction cross section becomes 
\begin{eqnarray} 
C_{\ext}=\cos^{2}\theta C_{\|}+\sin^{2}\theta C_{\perp}.
\label{eq:Cext0}
\end{eqnarray}

where $C_{\perp}$ and $C_{\|}$ denote the extinction of radiation for the cases in which electric field $\mathbf E$ is perpendicular and parallel to the symmetry axis $\bold a$, respectively. The effective cross-section for randomly-oriented spheroids can be written as 
%Since the original starlight is unpolarized, one can compute the total extinction cross section for a randomly oriented grain by integrating Equation (\ref{eq:Cext0}) over the isotropic distribution of $\theta$, i.e., $f_{\rm iso} d\theta \sim \sin\theta d\theta$. As a result,
\begin{eqnarray} 
C_{\ext}=\frac{1}{3}\left(2C_{\perp}+C_{\|}\right)
\label{eq:Cext}
\end{eqnarray}

For dust grains spinning around the principal axis $\bold a$, the polarization cross-sections are 
\begin{eqnarray}
\begin{aligned} 
C_{\pol}&=\frac{C_{\| ,\ext}-C_{\perp ,\ext}}{2} \hspace{0.9cm} {\rm for \hspace{0.2cm} prolates} \\
C_{\pol}&=C_{\| ,\ext}-C_{\perp ,\ext} \hspace{1.0cm} {\rm for \hspace{0.2cm} oblates}.
\end{aligned}
\label{eq:Cpol}
\end{eqnarray}

% ------------ extinction
\section{Extinction}
The extinction induced by randomly oriented grains in units of magnitude is defined by
\begin{eqnarray}
A({\lambda})&=&2.5{\rm log}_{10}\left(\frac{F_{\lambda}^{\rm obs}}{F_{\lambda}^{\star}}\right)=1.086\tau_{\lambda},\nonumber\\
&=&1.086\int \sum_{j=\rm carb,\rm sil}\sum_{i=0}^{N_{a}-1} C_{\ext}^{j}(a_{i})
n^{j}(a_{i})dz,
\label{eq:Aext}
\end{eqnarray}
where $F_{\lambda}^{\star}$ is the intrinsic flux from the star, $F_{\lambda}^{\rm obs}=F_{\lambda}^{\star}e^{-\tau{_\lambda}}$ is the observed flux, $\tau_{\lambda}$ is the optical depth, and the integration is performed along the line of sight.

To find the polarization by aligned grains, let us define an observer's reference system in which the line of sight is directed along the $z-$axis, and $x-$ and $y-$ axes constitute the sky plane. The polarization of starlight arising from the dichroic extinction by aligned grains in a cell of $dz$ is computed as
\begin{eqnarray}
dp({\lambda})=\frac{d\tau_{x}-d\tau_{y}}{2}=\sum_{i=0}^{N_{a}-1}\frac{1}{2}\left(C_{x}-C_{y}\right)
n(a_i)dz,
\label{eq:dplam}
\end{eqnarray}
where $N_{a}$ is the number of size bin, $n(a_{i})\equiv (dn/da)da$ is the number of grains of size $a_{i}$, $C_{x}$ and $C_{y}$ are the grain cross section along the $x-$ and $y-$ axes, respectively.

For the case of perfect internal alignment of grain symmetry axis $\ba_{1}$ with angular momentum $\bJ$, by transforming the grain coordinate system to the observer's reference system and taking corresponding weights, we obtain
\begin{eqnarray}
C_{x}&=&C_{\perp}-\frac{C_{\pol}}{2}\sin^{2}\beta,\\
C_{y}&=&C_{\perp}-\frac{C_{\pol}}{2}(2\cos^{2}\beta\cos^{2}\gamma+\sin^{2}\beta\sin^{2}\gamma),
\end{eqnarray}
where $\gamma$ is the angle between the magnetic field $\Bv$ and the sky plane and $\beta$ is the angle between $\bJ$ and $\Bv$.

The polarization efficiency then becomes
\begin{eqnarray}
C_{x}-C_{y}=C_{\pol}\frac{\left(3\cos^{2}\beta-1\right)}{2}\cos^{2}\gamma.
\label{eq:Cx-Cy1}
\end{eqnarray}
Taking the average of $C_{x}-C_{y}$ over the distribution of the alignment angle $\beta$, it yields
\begin{eqnarray}
C_{x}-C_{y}=C_{\pol} Q_{J}\cos^{2}\gamma,
\label{eq:Cx-Cy2}
\end{eqnarray}
where $Q_{J}=\langle G_{J}\rangle$ is the ensemble average of $G_{J}=\left(3\cos^{2}\beta-1\right)/2$ that describes the alignment of $\bJ$ and $\Bv$.

When the internal alignment is not perfect, following the similar procedure, we obtain
\begin{eqnarray}
C_{x}-C_{y}=C_{\pol}\langle 	G_{J}G_{X} \rangle \cos^{2}\gamma \equiv
C_{\pol}R\cos^{2}\gamma,
\label{eq:Cx-Cy3}
\end{eqnarray}
where $G_{X}=\left(3\cos^{2}\theta-1\right)/2$ with $\theta$ being the angle between $\ba_{1}$ and $\bJ$, and $R=\langle G_{J}G_{X}\rangle$ is the Rayleigh reduction factor (see also \citealt{1999MNRAS.305..615R}). The degree of internal alignment is described by $Q_{X}=\langle G_{X}\rangle$.

For a perpendicular magnetic field, i.e., $\Bv$ lies on the sky plane, Equation (\ref{eq:Cpol}) simply becomes $C_{x}-C_{y}=C_{\pol}R$. For an arbitrary magnetic field geometry, let $f=R\cos^{2}\gamma$ be the effective degree of grain alignment, which is a function of grain size $a$. Thus, in the following, $f(a)$ is referred as the alignment function.

Plugging the above equation into Equation (\ref{eq:dplam}) and integrating along the line of sight, we obtain
\begin{eqnarray} 
p({\lambda})=\int^{a_{\rm max}}_{a_{\rm align}} \sum_{j=\rm carb,sil} \frac{1}{2}C_{\pol}^{j}f^{j}(a)\frac{{\rm d}n}{{\rm d}a}{\rm d}a
\end{eqnarray}
where $f^{j}(a_i)$ is the effective degree of grain alignment for the grain specie $j$ of size $a_i$. 

It is more convenient to represent the extinction (polarization) through the extinction (polarization) cross section. Hence, the above equations can be rewritten as
\begin{eqnarray}
A({\lambda})&=&1.086~\sigma_{\ext}(\lambda)\times N_{\H},\\
p({\lambda})&=&\sigma_{\pol}(\lambda)\times N_{\H}
\label{eq:sigext}
\end{eqnarray}
where $N_{\H}(\cm^{-2})$ is the column density, and $\sigma_{\ext}$ and $\sigma_{\pol}$ in units of $\cm^{2} \H^{-1}$ are the dust extinction cross section and polarization cross section, respectively.

\end{document}